%% file: 2021_Prechelt_Tau_Near_Horizon.tex
\documentclass[aps,prd,twocolumn,groupedaddress,superscriptaddress,floatfix]{revtex4-2}

\usepackage{aas_macros}
\usepackage{amsmath}
\usepackage{lipsum}  
\usepackage{graphicx}
\usepackage{xcolor}
\usepackage{tabularx}
\usepackage{caption}
\usepackage{subcaption}
\usepackage[normalem]{ulem}

\graphicspath{{figures/}}
\newenvironment{conditions*}
  {\par\vspace{\abovedisplayskip}\noindent
   \tabularx{\columnwidth}{>{$}l<{$} @{${}={}$} >{\raggedright\arraybackslash}X}}
  {\endtabularx\par\vspace{\belowdisplayskip}}

\newcommand{\tapioca}{\texttt{tapioca}}

\begin{document}

\title{Analysis of a Tau Neutrino Origin for the Near-Horizon Air Shower
  Events Observed by the Fourth Flight of the Antarctic Impulsive Transient
  Antenna (ANITA)}

\include{anita_revtex_institutes}
\include{anita_revtex_authors}

\date{\today}

\begin{abstract}
We study in detail the sensitivity of the
  Antarctic Impulsive Transient Antenna (ANITA) to possible $\nu_\tau$ point source
  fluxes detected via $\tau$-lepton-induced air showers. This investigation
  is framed around the
  observation of four upward-going extensive air shower events very close to the horizon seen in ANITA-IV.
  We find that these four upgoing events are not observationally inconsistent with $\tau$-induced
  EASs from Earth-skimming $\nu_\tau$ both in their spectral properties as well as in their observed
  locations on the sky.
  These four events as well as the overall diffuse and point source exposure to Earth-skimming $\nu_\tau$ are also
  compared against published ultrahigh-energy neutrino limits from the Pierre Auger Observatory. While none of these four events occured at sky locations simultaneously visible by Auger, the
  implied fluence necessary for ANITA to observe these events is in strong tension with limits set by Auger across a wide
  range of energies and is additionally in tension with ANITA's Askaryan in-ice neutrino channel above~$10^{19}$~eV. We conclude
  by discussing some of the technical challenges with simulating and analyzing these near horizon events and the potential
  for future observatories to observe similar events.
\end{abstract}

\maketitle



\section{Introduction}

The fourth flight of ANITA (ANITA-IV) observed four below-horizon cosmic ray-like
events that have non-inverted polarity - 
a 3.2~$\sigma$ fluctuation if due to background~\cite{2021PhRvL.126g1103G}. Unlike the
steeply-upcoming~($\sim30^{\circ}$~below the radio horizon) anomalous events of
this type reported in two previous ANITA flights~\cite{2016PhRvL.117g1101G,
  2018PhRvD..98b2001G}, all of the ANITA-IV anomalous events are observed at angles
close to the horizon ($\lesssim 1^{\circ}$ below the horizon).

A  Standard Model explanation originally proposed for the steeply upcoming events
 from the first and third ANITA
flights (ANITA-I and ANITA-III, respectively) was skimming $\nu_\tau$ interactions in the Earth
producing $\tau$ leptons that escape into the atmosphere, subsequently decaying
and producing an upgoing extensive air shower~(EAS). While this origin was
initially considered to be unlikely due to the attenuation of neutrinos across
the long chord lengths through Earth at these steep
angles, several analyses have studied the $\nu_{\tau}$-origin
hypothesis for these steeply upgoing events~\cite{2016PhRvL.117g1101G, Aartsen_2020}.

These analyses have studied two different astrophysical assumptions: (1) that
the events were due to a diffuse isotropic flux of ultra-high energy
(UHE) neutrinos; and (2) that the events were from transient UHE neutrino
point sources that were active or flaring during each flight.

Under the diffuse hypothesis for the ANITA-I \& ANITA-III anomalous events, a prior analysis
by the ANITA collaboration~\cite{2019PhRvD..99f3011R,2019PhRvD..99f3011R_Erratum} implied a diffuse neutrino flux limit that is in strong tension 
with the limits imposed by the
IceCube~\cite{2016PhRvL.117x1101A} and Pierre Auger~\cite{2015PhRvD..91i2008A}
observatories~(Auger).

A preliminary follow-up analysis by the ANITA collaboration estimated the sensitivity of ANITA to $\nu_\tau$
point sources in the direction of the ANITA-I
and ANITA-III anomalous events to investigate the possibility that a point-like neutrino source could be
responsible for these events.
This analysis bounded the \textit{instantaneous} point source effective area to
$\lesssim$~2.2~m$^2$ for the ANITA-I event and $\lesssim$~0.3~m$^2$ for the ANITA-III event~\cite{Wissel:2019ot}. These values are significantly smaller than
Auger's $\nu_\tau$ point source peak effective area to $\nu_{\tau}$ of
$1\times10^4$ to $3\times10^5$~m$^2$ for energies above
$10^{17}$~eV and are also in strong tension with point-like neutrino limits set by Auger~\cite{2019JCAP...11..004A}

A number of alternative hypothesis have been proposed to explain the ANITA-I and
ANITA-III anomalous events. These range from Beyond Standard Model~(BSM)
physics~\cite{anomalous_1,anomalous_2,anomalous_3,anomalous_4,anomalous_5,anomalous_6,anomalous_7,anomalous_8,anomalous_9}
to more mundane effects such as transition radiation of cosmic ray air showers
piercing the Antarctic ice sheet~\cite{2019PhRvL.123i1102D} and subsurface
reflections due to anomalous ice features~\cite{2020AnGla..61...92S} although
the latter has recently been experimentally constrained by the
ANITA collaboration~\cite{2020arXiv200913010S}.


  

ANITA's sensitivity to the $\tau$ EAS channel is highly directional
and is maximal near the horizon where it is orders of magnitude larger than for
the steeply upgoing angles of the ANITA-I and ANITA-III events. This opens the possibility
for an Earth-skimming $\nu_\tau$ explanation for the ANITA-IV events (that occur extremely
close to the horizon) while potentially significantly reducing the tension with limit set by Auger and IceCube. 

The detection of this new class of events,
not observed in previous ANITA flights, is consistent with the improvements in
ANITA-IV's sensitivity~\cite{2019PhRvD..99l2001G}. In this paper we update the previous ANITA $\nu_\tau$ EAS
sensitivity analysis~\cite{2019PhRvD..99f3011R,2019PhRvD..99f3011R_Erratum} to estimate the diffuse and
point source transient fluxes implied by this new class of events and compare them with the
limits imposed by other neutrino observatories.


The paper is organized as follows: in \S\ref{sec:A4_events} we summarize the
properties of the ANITA-IV anomalous events relevant to this analysis; in
\S\ref{sec:sensitivity}, we present the simulation framework used to estimate
ANITA's sensitivity to $\tau$ neutrinos, via in-air $\tau$ decays, including a
discussion of the models used for the air shower radio emission and detector
response. The results of this analysis are presented in \S\ref{sec:results} and
detailed discussions and conclusions are provided in \S\ref{sec:discussion} and
\S\ref{sec:conclusion}, respectively.

\section{ANITA-IV Anomalous Events}
\label{sec:A4_events}

While ANITA was originally designed to detect the Askaryan emission from in-ice
UHE neutrino interactions, ANITA is also sensitive to the geomagnetic radiation emitted by ultrahigh-energy cosmic ray (UHECR) induced extensive air showers (EAS) as they develop in the atmosphere.
As an extensive air shower evolves in the
presence of the Earth's magnetic field, charged particles in the shower are
accelerated via the Lorentz force, creating a time-varying transverse current
within the shower. This transverse current generates an impulsive electric field
whose polarization is transversely aligned with the orientation of the Earth's magnetic
field. Over Antarctica, the Earth's magnetic field is primarily vertical,
resulting in predominately horizontally-polarized emission from an EAS~\cite{PhysRevD.86.123007}.

Typical air shower events observed by ANITA are classified into two
categories: (1) \textit{direct} UHECR events that reconstruct above the radio horizon
(i.e. ANITA observes the emission directly from the shower at it develops in the atmosphere); and (2)
\textit{reflected} UHECR events where ANITA observes the radio emission from air
showers after the radio emission \textit{has reflected off the surface of Antarctica}
(these must therefore reconstruct \textit{below} the horizon).

Along with their reconstructed direction, events are also typically classified
as direct or reflected by the polarity of the received electric field.
For a unipolar waveform, the polarity is determined by the sign~($\pm$) of the impulse.
For bipolar waveforms, the polarity is typically indicated by the order of the two primary poles (i.e. $+,-$ or $-,+$). Due to the Fresnel
reflection coefficient at the air-ice boundary, reflected EAS signals have a
completely inverted polarity with respect to the signals observed directly from
an EAS without reflection~\cite{Schoorlemmer:2015afa}. \textit{Polarity}, which
is related to the \textit{sign} of the electric field impulse, is distinct from
\textit{polarization}, which describes the \textit{geometric orientation of the
  electric field} and is used to separate EAS events from in-ice
Askaryan neutrino events. Over its four flights, ANITA has observed seven direct events
and 64 reflected UHECR
events~\cite{2010PhRvL.105o1101H,Schoorlemmer:2015afa,2021PhRvL.126g1103G}.

ANITA-IV also observed four extensive air shower-like events that have the same
polarity as the direct events (i.e. non-inverted $\implies$ non-reflected), but
reconstruct below the horizon.
These four events therefore appear to be upward-going air showers emerging from
the surface of the Earth, but unlike the ANITA-I and ANITA-III anomalous events,
the ANITA-IV events reconstruct near to, but below the horizon
($\lesssim1^{\circ}$)~\cite{2021PhRvL.126g1103G}. As shown in
Table~\ref{tab:A4_anomalous_events}, ANITA's angular uncertainty for these
events is $\sim0.2^{\circ}$, placing these events typically $\gtrsim1\sigma$ to $4\sigma$ below the horizon.

\begin{figure}
  \centering
  \includegraphics[width=\columnwidth]{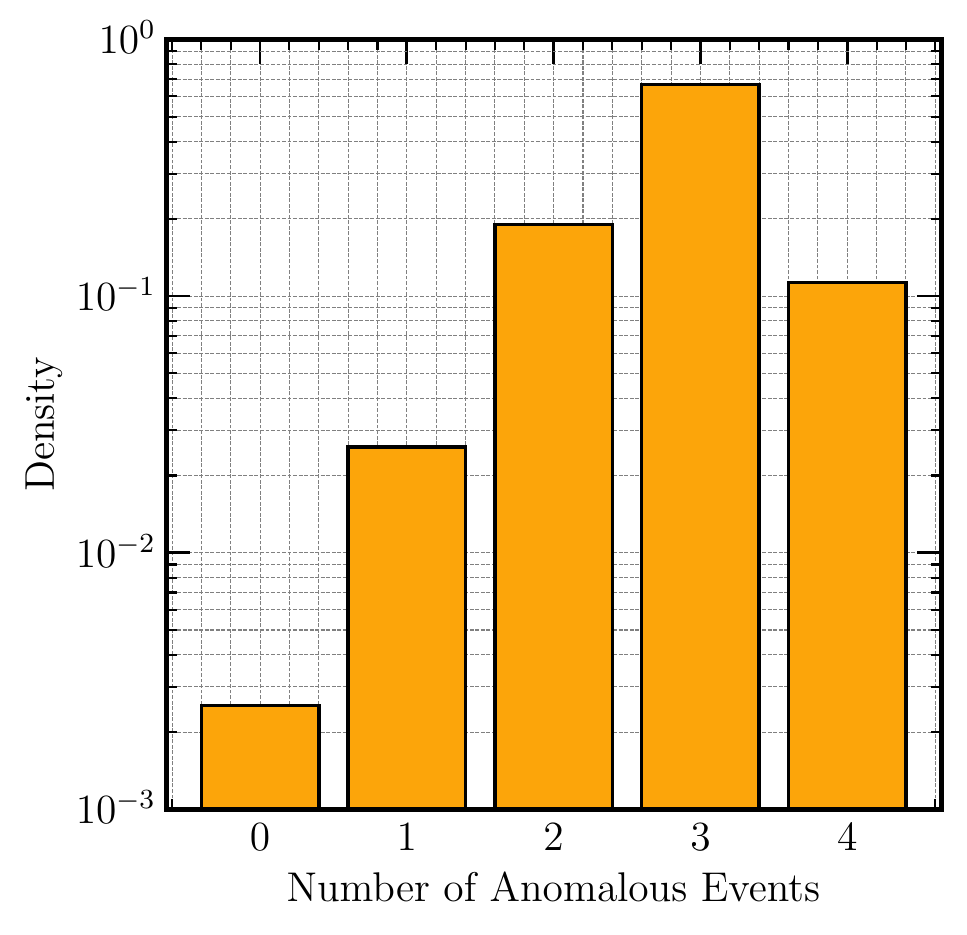}
  \caption{The probability density distribution (i.e. normalized event
    density) for the true number of EAS signal events observed by ANITA-IV
    under the assumptions of the toy Monte Carlo simulations described
    in~\cite{2021PhRvL.126g1103G}}.
  \label{fig:true_number_of_events}
\end{figure}

The significance of finding 4 events with a polarity inconsistent with their
geometry out of the 27 air shower events with a well-determined polarity is
estimated to be greater than 3\,$\sigma$, when considering the possibilities
that: the events could be an anthropogenic background; that there might be an
error in the reconstructed arrival direction; and that the polarity might be
misidentified~\cite{2021PhRvL.126g1103G}. The probability distribution of the
true number of non-inverted EAS-like events from one of the two Monte Carlo
simulations used to evaluate the above significance is shown in
Table~\ref{fig:true_number_of_events}. The most probable number of true signal
events, under the above background possibilities, is three with total probability density of 0.7 (against the probability of observing 0, 1, 2, or 4 signal events).

While the significance of the observed anomalies do not clearly distinguish
these events from possible backgrounds, in this study we consider the
hypothesis that these events may be due to a tau neutrino interaction inside the
Earth that creates an exiting $\tau$ lepton and analyze these events under this hypothesis.


The observed event parameters for the ANITA-IV
anomalous events under the hypothesis of a tau-neutrino origin are
shown in Table~\ref{tab:A4_anomalous_events} where $\theta$ is the elevation
angle of the observed radio-frequency direction and $\theta_H$ is the elevation
angle of the radio horizon.

\begin{table}
  \caption{\label{tab:A4_anomalous_events} The time of each observed event along with the 
  reconstructed elevation angle of the shower, $\theta$, below the radio horizon, $\theta_H$. All
  events are observed within $1^\circ$ of the radio horizon and are all greater than $1\sigma$ from the horizon.}
\begin{ruledtabular}
\begin{tabular}{r c c}
Event & Time (ISO 8601) & $(\theta - \theta_H)$ (deg.) \\
  \hline
  4098827  & 2016-12-03T10:03:27Z & -0.25$\pm$0.21$^{\circ}$  \\
  19848917 & 2016-12-08T11:44:54Z & -0.65$\pm$0.20$^{\circ}$  \\
  50549772  & 2016-12-16T15:03:19Z & -0.81$\pm$0.20$^{\circ}$  \\
  72164985  & 2016-12-22T06:28:14Z & -0.19$\pm$0.10$^{\circ}$  \\
\end{tabular}

\end{ruledtabular}
\end{table}

\section{ANITA's Sensitivity to Tau Neutrinos via Extensive Air Showers}

\label{sec:sensitivity}

In this section we estimate the sensitivity of ANITA to tau neutrinos via
the extensive air shower channel for both diffuse and point source fluxes. A
flux density of tau neutrinos $F(t, E_\nu, \hat{r})$ arriving on Earth (i.e. the
number of events per unit area per unit solid angle per unit time per unit
energy) depends on sky direction $\hat{r}$, and varies with neutrino energy
$E_\nu$ and time $t$. The differential contribution to the number of events
$N_{obs}$ detected with a given observatory is shown in Equation \ref{eq:event_rate_general}.


\begin{equation}
\begin{split}
    \frac{dN_{obs}}{dt \ dE_\nu \ d\Omega \ dA } =  ({\hat{r}_\nu}\cdot\hat{x}_E) \ & \Theta(\hat{r}\cdot\hat{x}_E)  \\ & F(t, E_\nu, \hat{r}) \ P_{obs}(t, E_\nu, \hat{r}, \vec{x}_E).
    \label{eq:event_rate_general}
  \end{split}
  \end{equation}
where:
  \begin{conditions*}
    dN_{obs} & the differential number of observed events, \\
    dA & the differential area we consider, \\
    t & the observation time, \\
    dt & the differential observation time interval, \\
    E_\nu & the neutrino energy, \\
    dE_\nu & the differential neutrino energy band, \\
    \hat{r}_\nu & the vector point towards the neutrino source, \\
    \vec{x}_E & the location of the differential area on the surface of the Earth, \\
    \Theta & the Heaviside step-function, \\
    F(t, E_\nu, \hat{r}) & the incident flux of tau neutrinos, \\
    P_{obs}(t, E_\nu, \hat{r}, \vec{x}_E) & the probability that this event is detected by this observatory.
    \end{conditions*}

The behavior specific to the observatory is encoded in the function
$P_{obs}(t, E_\nu, \vec{x}_E, \hat{r})$, which is the probability that a tau
neutrino with energy $E_\nu$ coming from sky direction $\hat{r}$ whose axis of
propagation intersects a point $\vec{x}_E$ on the surface of integration $A$
(the surface of the Earth for ANITA) at time $t$ is observed. The dot product
$\hat{r}\cdot\hat{x}_E$ accounts for the projected area element in the direction
of $\hat{r}$ and $\Theta(\hat{r}\cdot\hat{x}_E)$ accounts for the fact that we
only consider neutrinos propagation axes that exit the Earth in the observing regions of interest.

A full description of the function $P_{obs}$ for ANITA is derived for a diffuse flux
in~\cite{2019PhRvD..99f3011R,2019PhRvD..99f3011R_Erratum} and is extended for neutrino point sources in Equation \ref{eq:anita_pobs}.
The probability of observation is decomposed into
a convolution of probabilities summarized as follows: 1) the probability, $P_{\mathrm{exit}}$, that a
$\nu_\tau$ with original energy $E_\nu$ undergoes a sequence of interactions inside the
Earth that results in a $\tau$ lepton leaving the Earth; 2) the probability, $P_{\mathrm{decay}}$,
that the $\tau$ lepton subsequently propagates
in the atmosphere and decays before reaching ANITA~\cite{Wissel:2019ot}; 3) that the $\tau$ decay creates
a shower with sufficient energy to be detectable by ANITA and that the decay point is far away enough from ANITA that the shower can fully develop before passing ANITA; and 4) the probability,
$P_{\mathrm{trig}}$, that the radio emission 
from this particular shower is sufficiently strong enough to trigger ANITA (and be detected)~\cite{2012APh....35..325A,2019PhRvD..99f3011R,2019PhRvD..99f3011R_Erratum}.
All of 
these factors are combined later in this article to calculate ANITA's \textit{effective area} to $\tau$-induced
extensive air showers (Equation~\ref{eq:anita_pobs}) given a $\nu_\tau$ of energy $E_\nu$ coming from
direction $\hat{r}_\nu$ at time $t$ with the $\tau$ exiting the earth at $\vec{x}_E$. 



\begin{equation}
\begin{aligned}
 P_{obs}(t, E_{\nu}, \hat{r}, \vec{x}_E) = & \int dE_{\tau}\  P_{exit}\left(E_{\tau}\ |\ E_{\nu}, \theta_{em}  \right)\\[3pt]
 & \int ds_{decay}\ P_{decay}\left(s_{decay}\ |\ E_{\tau} \right)  \\[3pt]
 & \int dE_{EAS}\ P_{EAS}\left( E_{EAS}\ |\ E_{\tau} \right)  \\[3pt]
 & \int d\mathcal{E}\ P_{\mathcal{E}}\left(\mathcal{E}\ |\ E_{EAS},s_{decay},\hat{r}_{\nu},\right)\\[3pt]
 & \quad\quad\ \ P_{trig} \left(\vec{x}_{\mathrm{ANITA}}\ |\ \mathcal{E},s_{decay},\hat{r}_{\nu}\right)
    \label{eq:anita_pobs}
    \end{aligned}
  \end{equation}
where:
  \begin{conditions*}
    E_\tau & the energy of the exiting $\tau$, \\
    \theta_{em} & the emergence angle of $\tau$ at the surface, \\
    s_{decay} & the decay length of the $\tau$, \\
    E_{EAS} & the energy of the extensive air shower, \\
    \mathcal{E} & the electric field at the payload (V/m), \\
    \vec{x}_{\mathrm{ANITA}} & the location of ANITA, \\
    \hat{r}_{\nu} & the incident direction of the neutrino and $\tau$, \\
    \end{conditions*}

In \S\ref{sec:models} we provide some details of recent updates to the
simulation components that were previously used to estimate the diffuse flux sensitivity of ANITA-I and
ANITA-III~\cite{2019PhRvD..99f3011R, Wissel:2019ot}. Namely, the sampling of tau
decay modes, the look-up tables of electric fields produced with ZHAireS, and
significant refinements to the ANITA-IV detector model. \S\ref{sec:acceptance} reviews the
detector acceptance calculation used for estimating the sensitivity to a diffuse
neutrino flux and \S\ref{sec:pt_src_eff_area} develops the detector point source
effective area formalism used to estimate ANITA's sensitivity to a point source flux.


\subsection{Models}
\label{sec:models}

The models used to evaluate ANITA's acceptance to tau neutrinos via the extensive
air shower channel were first developed in~\cite{2019PhRvD..99f3011R, Wissel:2019ot}.
In this section, we discuss the improvements that have been made to these models in
this latest simulation to improve fidelity. For more detail and verification on the underlying models, we point
the reader to the aforementioned references.

\subsubsection{Tau Decays}

\begin{figure}
  \centering
  \includegraphics[width=\columnwidth]{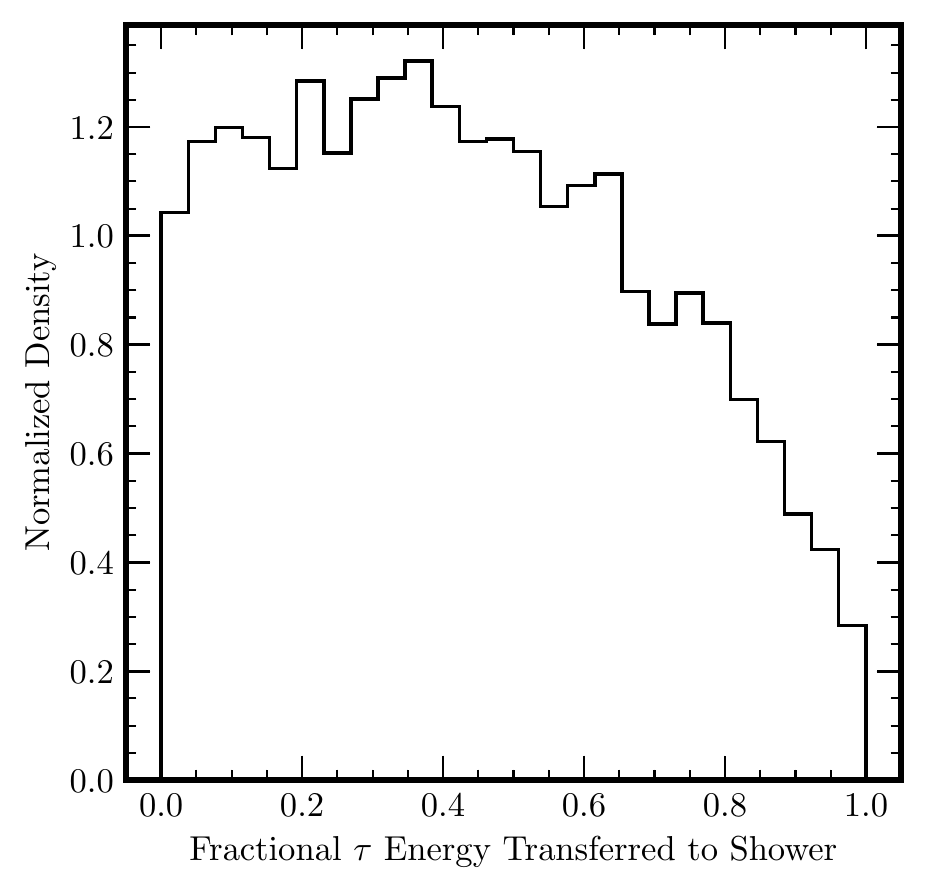}
  \caption{The relative fraction of the tau energy transferred to the shower in decays
    of ultrahigh-energy negative helicity $\tau$-leptons
    simulated using PYTHIA~8.244~\cite{2015CoPhC.191..159S,2019PhRvD..99f3011R,2019PhRvD..99f3011R_Erratum}.}
  \label{fig:tauola_samples}
\end{figure}

In the ANITA tau neutrino acceptance bounds
of~\cite{2019PhRvD..99f3011R,2019PhRvD..99f3011R_Erratum}, a single $\tau$ decay
was used ($\tau^-\to\pi^-\pi_0\nu_\tau$) with 99\% of the energy going to
showering particles to seed the air shower simulations. For this analysis and
in~\cite{Wissel:2019ot}, a large sample of $\tau^-$ decays with negative
helicity were sampled using PYTHIA 8.244~\cite{2015CoPhC.191..159S}. For each
sample, we estimated the total fraction of initial energy that goes into the EAS
by excluding neutrinos and muons from the secondary particles. The resulting
fraction of the original $\tau$ energy transferred to the air shower is shown in
Figure \ref{fig:tauola_samples}.

\subsubsection{Air Shower Electric Field Model}
\label{sec:field_models}

\begin{figure}
  \centering
  \includegraphics[width=\columnwidth]{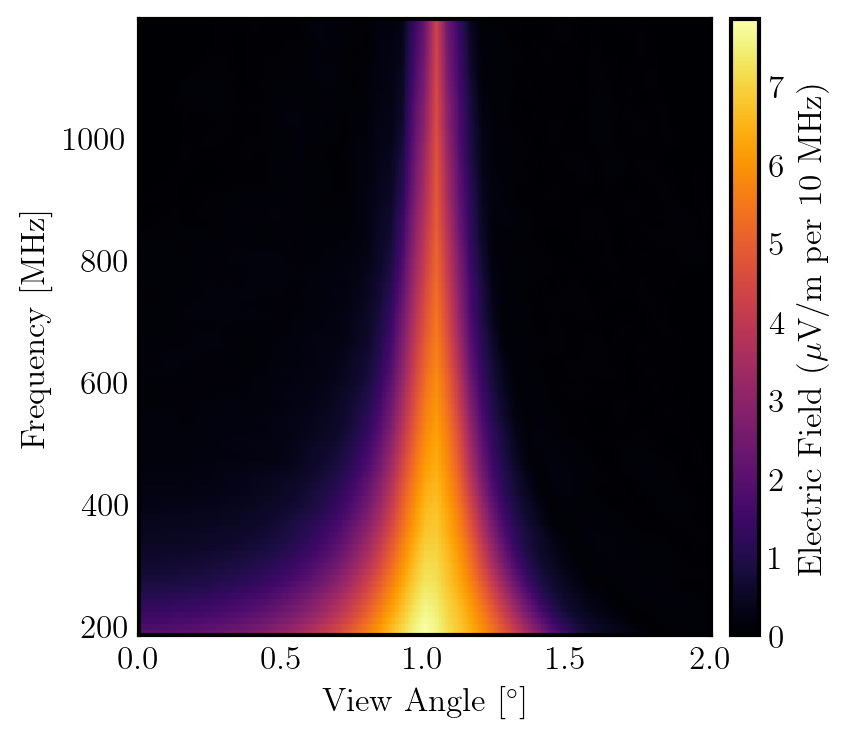}
  \caption{The electric field strength as a function of frequency and view angle with respect
  to the shower axis
    for a specific \textit{slice} through the 4-D lookup table used to evaluate
    the electric field ($\theta_{\mathrm{em}} = 5^{\circ}$,
    $h_{\mathrm{decay}} = 2$~km).}
  \label{fig:electric_field_spectrum}
\end{figure}

\begin{figure}
  \centering
  \includegraphics[width=\columnwidth]{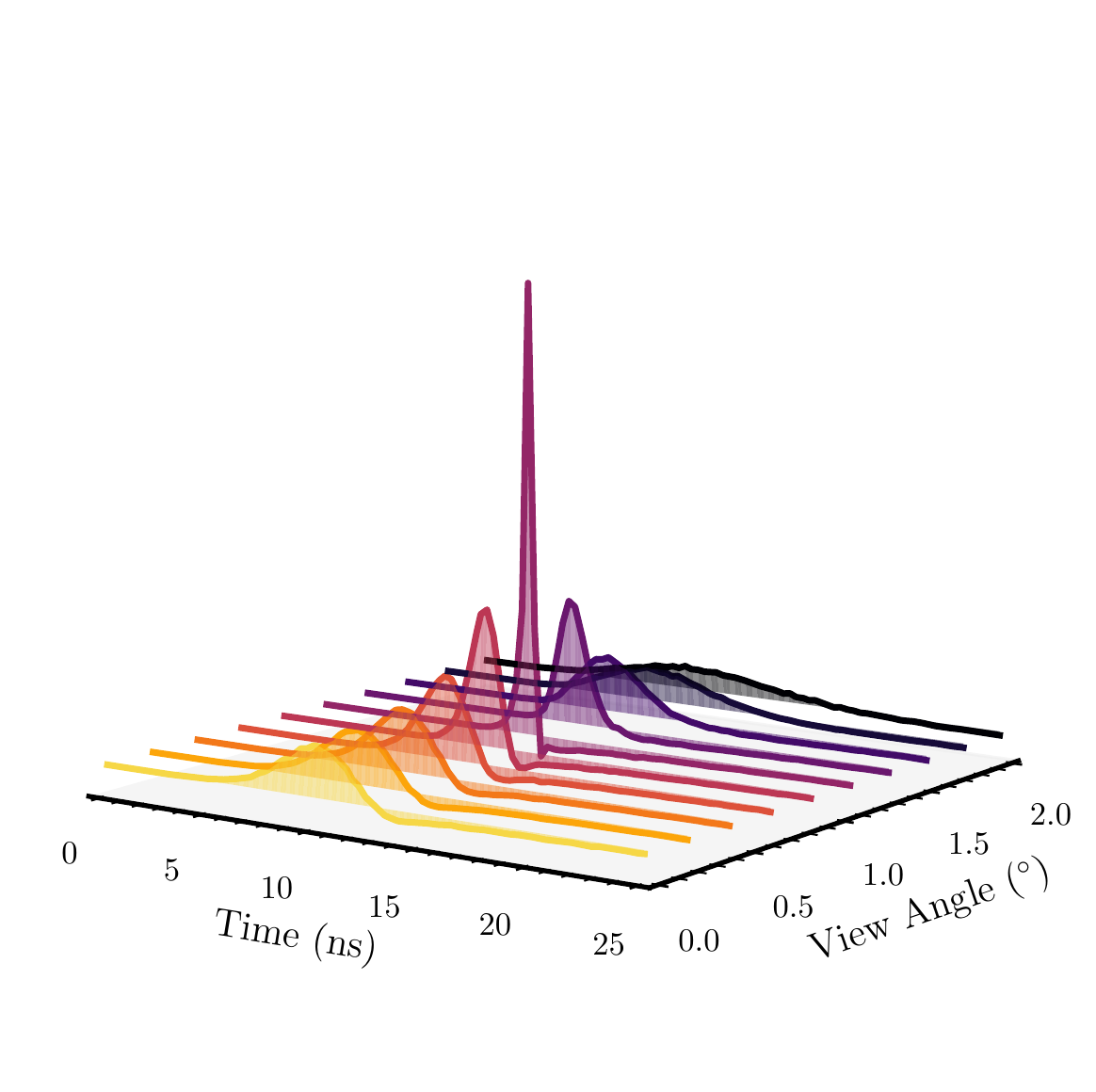}
  \caption{ZHAireS-simulated electric field waveforms for various view angles
    for a specific \textit{slice} through the 4-D lookup table used to evaluate
    the electric field ($\theta_{\mathrm{em}} = 5^{\circ}$,
    $h_{\mathrm{decay}} = 2$~km).}
  \label{fig:electric_field_waveforms}
\end{figure}

The previous ANITA-I \& ANITA-III $\nu_\tau$ analysis used a simplified electric
field model accounting only for the \textit{peak} value of the electric field
based on a large sample of ZHAireS~\cite{2012APh....35..325A} simulations. These
were run at a full range of decay altitude ($0 - 9$~km in 1~km steps), emergence
angles ($1^{\circ} - 35^{\circ} $), and off-axis view angles, $\psi$
($0^\circ - 3^\circ$) completely covering the amplitude range detectable by
ANITA. For this new work, this model was significantly improved by storing the
complete time-domain electric field waveform (see
Figure~\ref{fig:electric_field_spectrum} and
Figure~\ref{fig:electric_field_waveforms} for an example). This EAS shower
library was then used to produce a pair of 4D lookup-table (LUT) of electric
field in terms of $(h_{decay}, \theta_{em}, \phi, f)$ or
$(h_{decay}, \theta_{em}, \phi, t)$ where $\theta_{em}$ is the emergence angle of the
$\tau$ at the exit location on the surface. All showers were simulated with a
shower energy of 100~PeV. The electric field amplitude
for EAS scales very close to linearly with shower energy across the energy range of
interest\cite{Schoorlemmer:2015afa,2012APh....35..325A}.


$h_{decay}$, $\theta_z$, and $\phi$ are known a priori based
upon the geometric properties of the specific simulation or Monte Carlo trial. These parameters are then used to
perform a 4D quad-linear interpolation into the lookup-table to estimate the
electric field at the payload. This field was
then used as the primary input to the antenna and detector model with
appropriate scaling to account for the specific shower energy that was being simulated.

\subsubsection{Detector Model}
\label{sec:detector_model}

The previous diffuse $\nu_\tau$ analysis used a detector model implemented
in the frequency domain and assumed a constant
frequency-independent boresight gain for ANITA's quad-ridged horn
antennas and ignored any off-axis effects due to the antenna beamwidth~\cite{2019PhRvD..99f3011R}. For this work, we have developed a completely new \textit{time-domain} detector model based on a recent measurement of the impulse responses of all 96 channels in ANITA-IV~\cite{2021PhRvL.126g1103G} and flight-data-driven noise model, along with additional detector model improvements.


This new \textit{time-domain} detector model directly uses
in-flight data and post-flight calibration measurements and as such is expected
to have much higher fidelity than the simple assumptions used in the frequency
domain model of~\cite{2019PhRvD..99f3011R}.


During the ANITA-IV EAS and neutrino analysis presented in~\cite{2021PhRvL.126g1103G}, a detailed
calibration campaign was performed to measure the total time-domain transfer function of
every channel and configuration of the ANITA-IV payload. ANITA-IV had 96 channels (48 antennas each
with two polarizations per antenna), however the payload also included dynamic notch filters on each channel,
that were reconfigured during the flight, to combat radio-frequency interference. These notch filters
can significantly change the amplitude and phase response of each individual channel; 6 different filter
configurations were used throughout the flight for a total of $\sim600$ independent impulse responses.

\begin{figure}
    \includegraphics[width=\columnwidth]{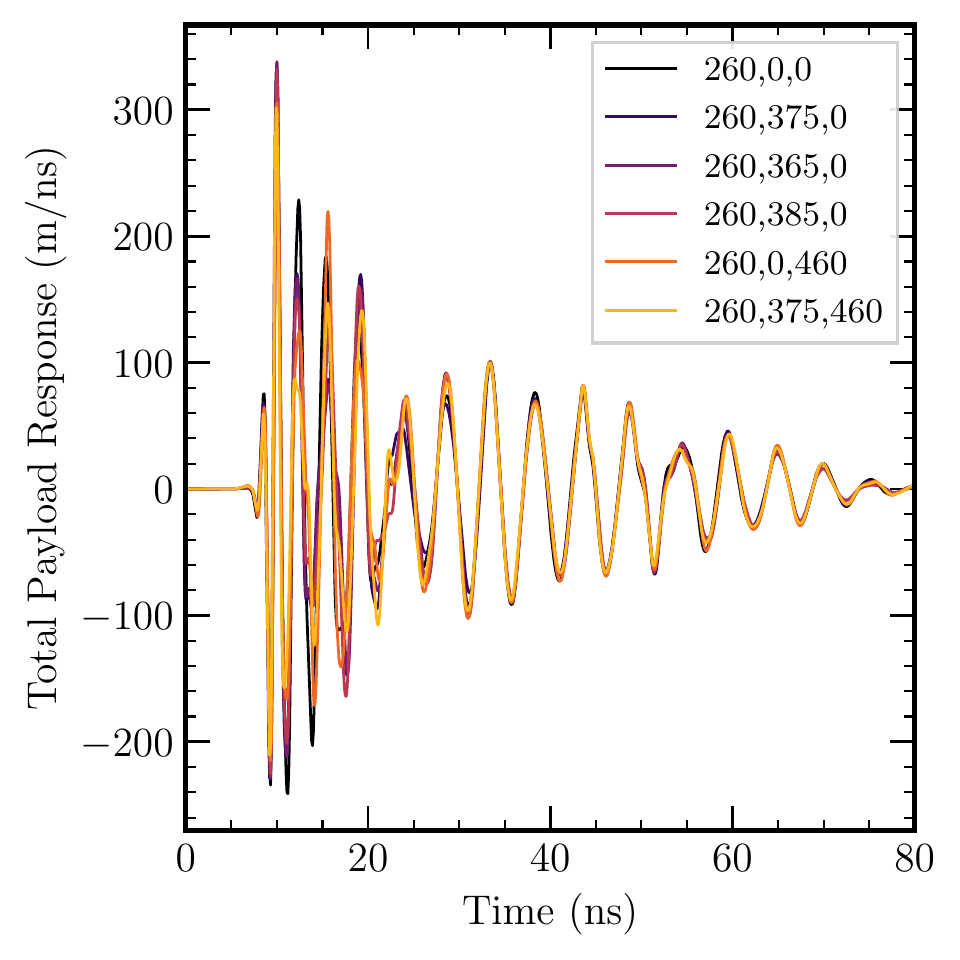}
    \caption{The \textit{average} time-domain full signal chain transfer function for
    each notch filter configuration used in the ANITA-IV flight. The legend indicates the frequencies
    (in MHz) that each of the three filter notches were programmed to for each configuration.}
    \label{fig:time_domain_responses}
  \end{figure}

The
payload-wide average of these responses for each configuration is shown in Figure~\ref{fig:time_domain_responses}. 
These measurements provide the total time-domain transfer function of each channel
(including antennas, front-end LNAs, bandpass filters, notch filters, and second-stage signal chain).
Since the transfer-function, $\bar{h}(t)$, is a complete representation of the ANITA signal chain from the
antenna to the digitizer, the incident electric field can be immediately converted
into a voltage via

\begin{equation}
    V(t) = \bar{h}(t) * \mathcal{E}(t)
    \label{eq:transfer_function_convolution}
\end{equation}
where:
\begin{conditions*}
    V(t) & the time-domain voltage measured by ANITA (in V), \\
    \bar{h}(t) & the time-domain transfer function (in m/ns), \\
    \mathcal{E}(t) & the time-domain incident electric field (in V/m), \\
    * & represents linear convolution,
    \end{conditions*}

To calculate the observed waveform for a given channel,
we convolve the time-domain
electric field from the 4D ZHAireS interpolation (described in \S\ref{sec:field_models}) with the specific time-domain response
for that channel.

\begin{figure}
    \includegraphics[width=\columnwidth]{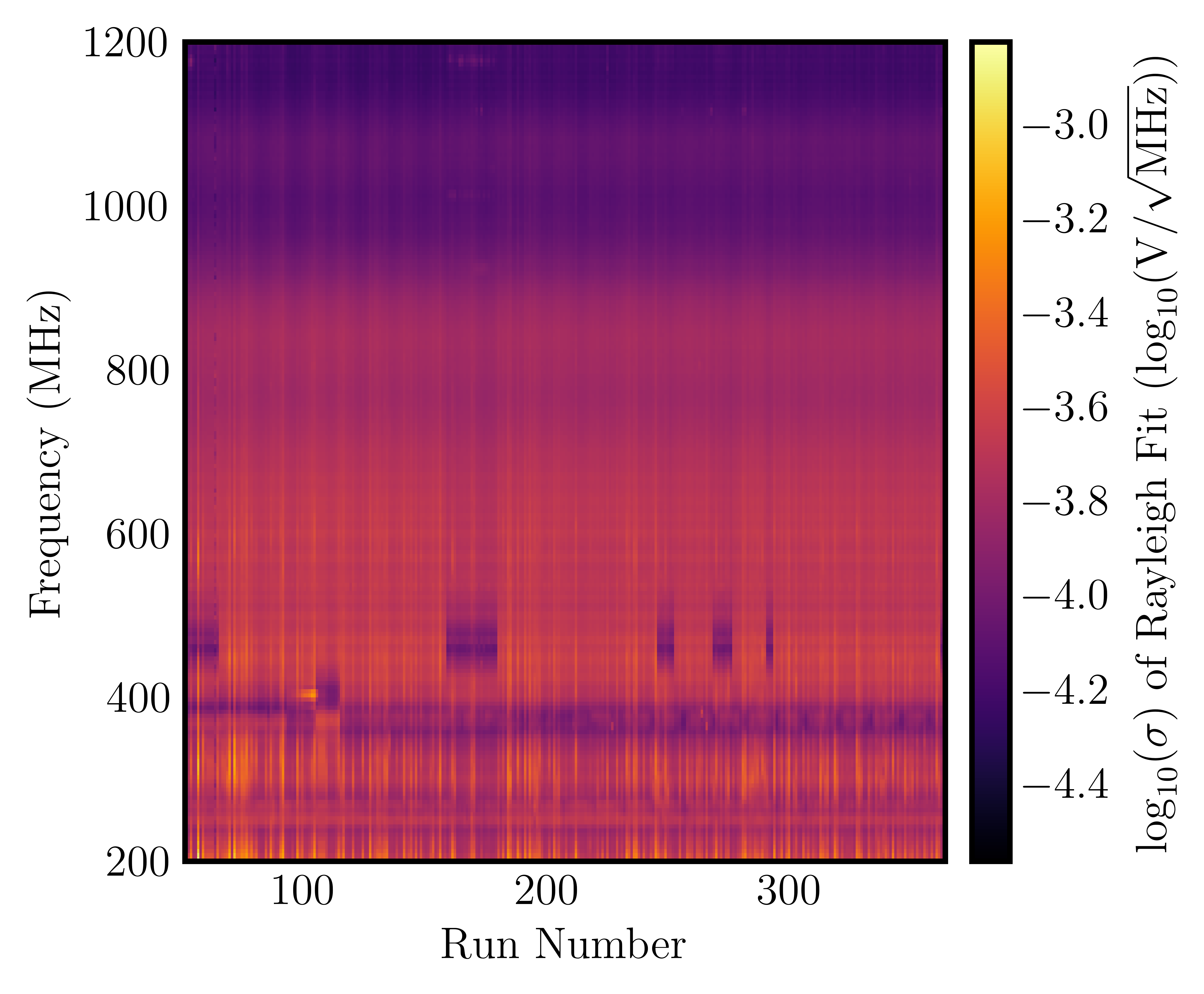}
    \caption{The average per-channel integrated noise power for each run of the
      ANITA-IV flight shown as the logarithm of the scale parameter, $\sigma$,
      of a Rayleigh distribution fit to the distribution of noise samples in a
      given frequency bin over a given run. The changing center frequency of the
      notch filters is clearly evident as changes in the location of the
      ``horizontal" stripes in this figure.}
    \label{fig:noise_over_flight}
  \end{figure}

  For the time-domain model, we replace the analytic noise generation model used
  in the previous ANITA $\nu_\tau$ analysis with a model derived from actual
  GPS-triggered noise waveforms recorded during the ANITA-IV flight. We
  extracted a large sample of so called \textit{``minimum-bias''} events from
  each of the $\sim$300 runs in the ANITA-IV flight dataset.
Since the bin-by-bin amplitude of random phasor noise in the frequency domain is expected to follow
a Rayleigh distribution, we fit a 
Rayleigh probability density function to each 5~MHz bin in the distribution of amplitude spectral densities using a 
maximum-likelihood estimator (see \cite{Cremonesi:2019zzc} for more details on this process).
The amplitude spectral density of background
noise evolved significantly throughout the flight (Figure~\ref{fig:noise_over_flight}), so this
fitting process is done on a run-by-run basis.

Given a simulated EAS signal waveform calculated using Equation~\ref{eq:transfer_function_convolution}
and a random noise waveform sampled from the fitted Rayleigh distributions, the signal-to-noise ratio (SNR), used
in our trigger calculation, is calculated with Equation~\ref{eq:time_domain_snr}.

\begin{equation}
  \begin{split}
    \mathrm{SNR} = & \frac{\max{V(t)} - \min{V(t)}}{2\sigma}\\ = &
    \frac{\max{V(t)} - \min{V(t)}}{2\sqrt{\sum_t V_{n}(t)^2}}
    \label{eq:time_domain_snr}
    \end{split}
\end{equation}
where:
  \begin{conditions*}
    SNR & signal-to-noise ratio, \\
    V(t) & the time-domain signal voltage, \\
    \sigma & the standard deviation of the noise waveform, \\
    V_{n}(t) & the simulated noise waveform \\
\end{conditions*}



\paragraph{Trigger Model}

\begin{figure}
\includegraphics[width=1\columnwidth]{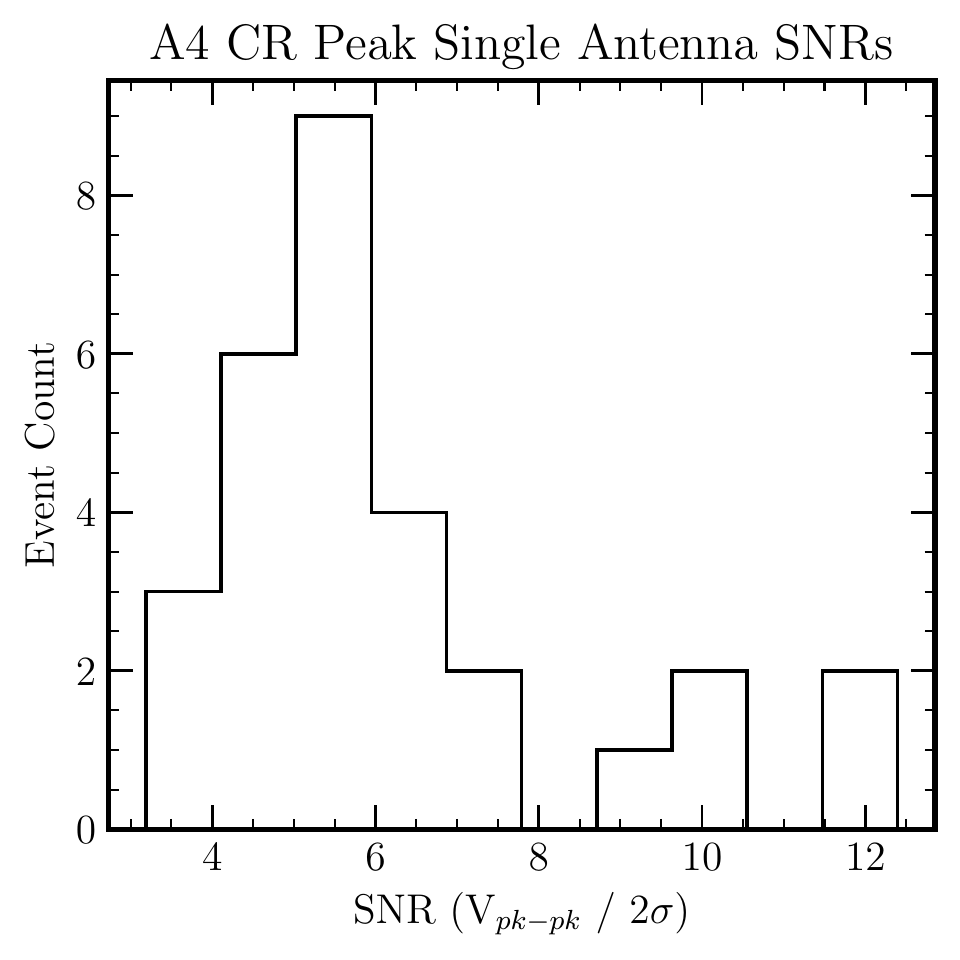}
  \caption{The peak single-antenna signal-to-noise ratio (SNR) for each of the 28 cosmic
    ray observed by ANITA-IV. $\sigma$ is the RMS of thermal noise measured in the corresponding channel}
     \label{fig:a4_cr_snrs}
\end{figure}

Given the observed SNR of a particular simulated event, we model ANITA's trigger using a Heaviside step-function where $P_{\mathrm{trig}} = 100\%$ for any trial with $\mathrm{SNR} > \mathrm{SNR}_{\mathrm{trig}}$ and $P_{\mathrm{trig}} = 0\%$ otherwise. The threshold signal-to-noise ratio, SNR$_{\mathrm{trig}}$, is chosen as the minimum value of the distribution of SNRs of the total population of EAS observed by ANITA-IV during post-flight analysis (Figure \ref{fig:a4_cr_snrs}). This was chosen to emulate both the hardware trigger as well as the analysis efficiency and cuts employed by ANITA-IV to isolate UHECR signal events.

In the following sections, we present the formalism used to calculate ANITA-IV's effective area and exposure
to $\tau$-induced EAS and present the results of these simulations for both diffuse and point-source $\nu_\tau$ fluxes.

\subsection{Acceptance to a Diffuse and Isotropic Flux}
\label{sec:acceptance}

We model a diffuse, isotropic flux as independent of direction and time so that we may write $F(t, E_\nu, \hat{r}) \simeq F(E_\nu)$. Equation~\ref{eq:event_rate_general} then becomes
\begin{equation}
\begin{split}
    \frac{dN_{obs}}{dt \ dE_\nu \ dA \ d\Omega} = (\hat{r}_\star\cdot\hat{x}_E) \ & \Theta(\hat{r}_\star\cdot\hat{x}_E)\\ & F(E_\nu) \ P_{obs}(t, E_\nu, \hat{r}_\star, \vec{x}),
    \end{split}
\end{equation}
where $\hat{r}_\star$ is a fixed direction on the sky. Since a diffuse isotropic flux, $F(E_{\nu})$, does not explicitly depend on the differential area of integration $dA$ or projected differential solid angle $d\Omega$, we may write
\begin{equation}
    \frac{dN_{obs}}{dt \ dE_\nu } = F(E_\nu) \ 
    \langle A\Omega \rangle(t, E_\nu),
\end{equation}
where the acceptance $A\Omega(t, E_\nu)$, averaged over the sky, is given by 
\begin{equation}
  \begin{split}
     \langle A\Omega \rangle(t, E_\nu) = & \int_\Omega d\Omega \\ & \int_A dA  \ \hat{r}_\star \cdot\hat{x}_E \ \Theta(\hat{r}\cdot\hat{x}_E)\  \ P_{obs}(t, E_\nu, \hat{r}_\star, \vec{x}),
     \label{eq:diffuse_acceptance}
     \end{split}
\end{equation}

This is based on the formalism used in~\cite{2019PhRvD..99f3011R,2019PhRvD..99f3011R_Erratum} but we have rederived
it here with an explicit time dependence to enforce a clear connection with the point source effective
area, derived in the next section. 
While the flux is independent of time, ANITA's acceptance, $\langle A\Omega \rangle(t)$, still has an explicit time dependence due to variations in $dA$ and $P_{obs}$ throughout the $\sim30$-day ANITA flight. 

\subsection{Point Source Effective Area}
\label{sec:pt_src_eff_area}

\begin{figure}
  \centering
  \includegraphics[width=1\columnwidth]{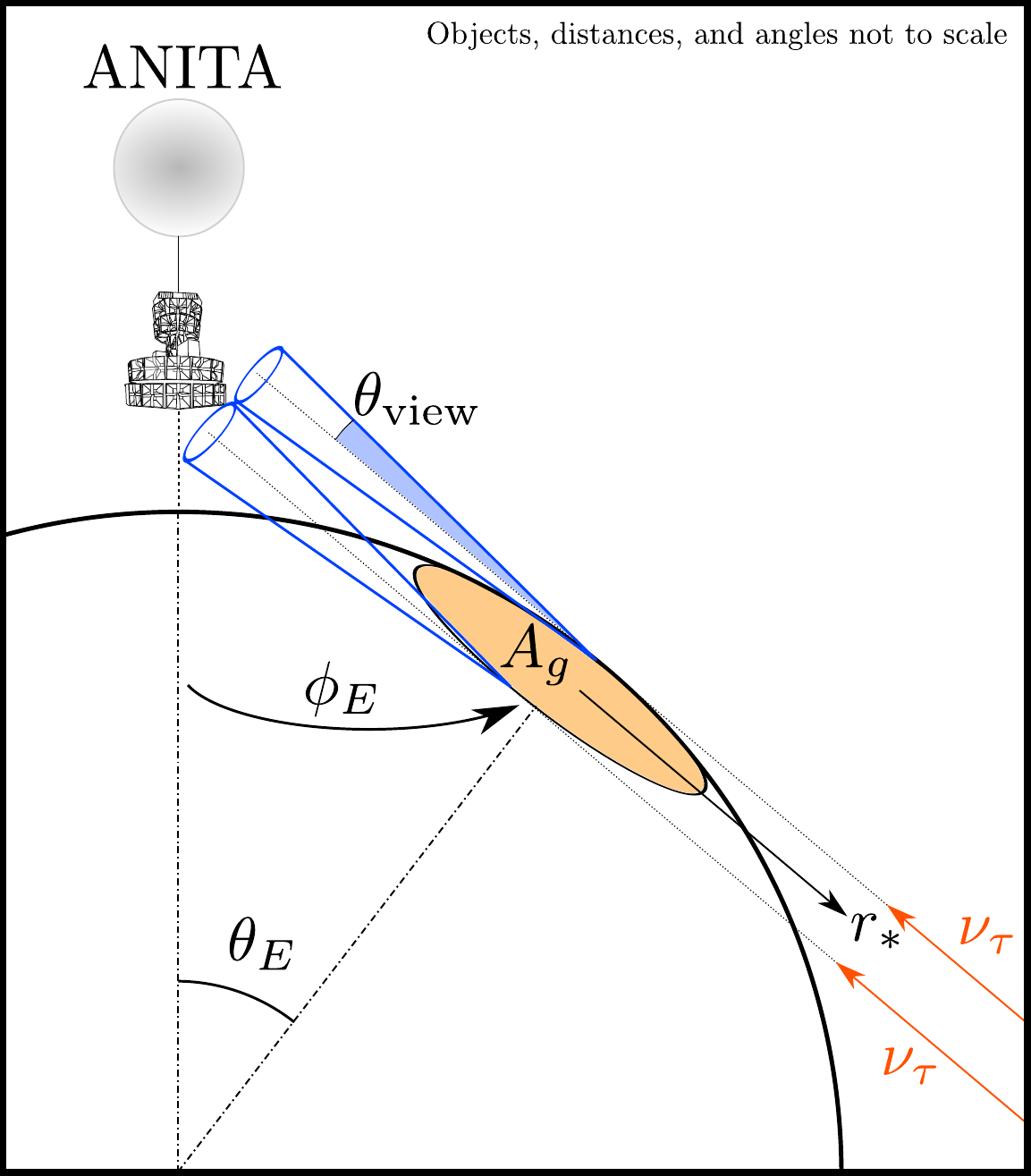}
  \caption{A diagram showing how this analysis defines the geometric area as the area 
  on the surface of the Earth which ANITA observes with a view angle from the surface 
  of less than $\theta_{{\mathrm{view, max}}}$.}
  \label{fig:geometric_area_diagram_a}
\end{figure}

We model a point source flux in a fixed sky direction $\hat{r}_\star$ as 
\begin{equation}
    S(t, E_\nu, \hat{r}_\star) = \int d\Omega \ \delta(\hat{r}-\hat{r}_\star) \ F(t, E_\nu, \hat{r}),
\end{equation}
where $\delta(\hat{r}-\hat{r}_\star)$ is a Dirac $\delta$-function on a
spherical surface with units of inverse steradians. Unlike a diffuse and isotropic
flux, point source flux is in units of number per
unit time per unit energy per unit area. Applying the integration over solid
angle in Equation~\ref{eq:event_rate_general}, the event rate for the point
source is given by
\begin{equation}
  \begin{split}
    \frac{dN_{obs}}{dt \ dE_\nu \ dA} = (\hat{r}_\star\cdot\hat{x}_E) \ & \Theta(\hat{r}_\star\cdot\hat{x}_E)\ \\& S(t, E_\nu, \hat{r}_\star) \ P_{obs}(t, E_\nu, \hat{r}_\star, \vec{x}).
    \end{split}
\end{equation}
Since the point source flux $S(t, E_\nu, \hat{r}_\star)$ also does not explicitly depend on the differential area of integration $dA$ used for estimating the event rate, we may write
\begin{equation}
    \frac{dN_{obs}}{dt \ dE_\nu } = S(t, E_\nu, \hat{r}_\star)
    \langle A \rangle(t, E_\nu, \hat{r}_\star),
\end{equation}
where the point source effective area $A(t, E_\nu, \hat{r}_\star)$ is given by 
\begin{equation}
  \begin{split}
     A(t, E_\nu, \hat{r}_\star) = \int_{A_g} dA_g \ (\hat{r}_\star\cdot\hat{x}_E) \ & \Theta(\hat{r}_\star\cdot\hat{x}_E) \\ & P_{obs}(t, E_\nu, \hat{r}_\star, \vec{x}),
     \label{eq:effective_area}
     \end{split}
\end{equation}

\noindent where $A_g$ is the geometric area on the surface of the Earth that we integrate
over (yellow spherical patch in Figure \ref{fig:geometric_area_diagram_a}). For ANITA's $\nu_\tau$ effective area, we define $A_g$ as the ellipsoidal
area on the surface from which the off-axis view angle at the surface,
$\theta_{\mathrm{view,exit}}$, is less than some predefined
$\theta_{\mathrm{max}}$. In most cases, $\theta_{\mathrm{max}} \sim 1-2^\circ$
(Figure \ref{fig:geometric_area_diagram_a}.


\subsection{\tapioca}
\label{sec:tapioca}

To simulate ANITA's sensitivity to both diffuse and point sources of $\tau$
neutrinos using the formalism in \S\ref{sec:pt_src_eff_area}, we have developed
a new $\nu_\tau$ simulation code, the \textbf{Ta}u \textbf{P}o\textbf{i}nt
S\textbf{o}urce \textbf{Ca}lculator, or \tapioca. \tapioca~is publicly available and released under an open-source copy-left
license. 

\begin{figure*}
  \includegraphics[width=1\textwidth]{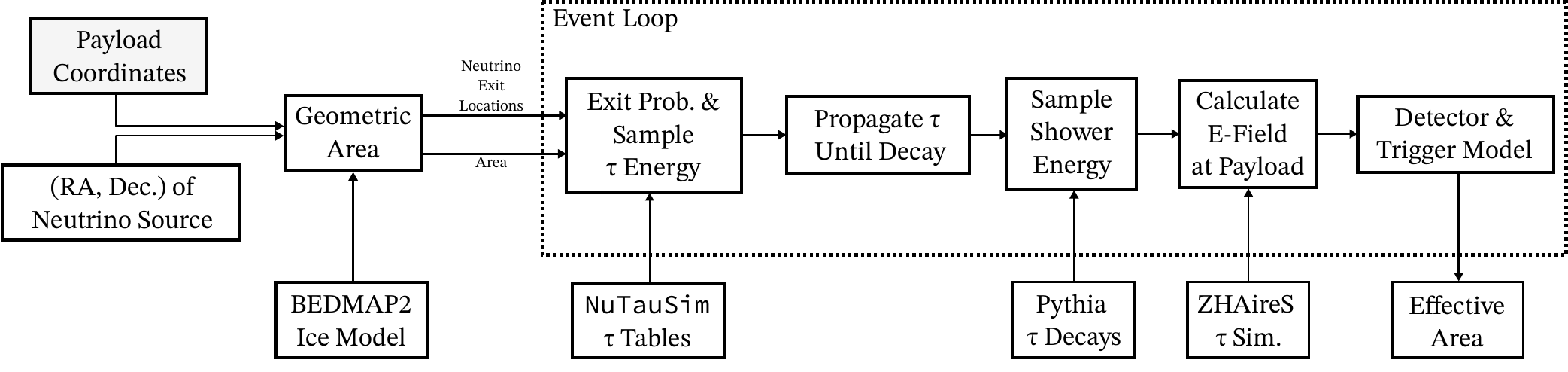}
  \caption{A flowchart showing the top-level logic and event loop of the
    \tapioca~simulation code with the external data sources used at each stage. This
    logic is described in detail in Section \ref{sec:tapioca} of the text.
    \label{fig:eventloop}}
\end{figure*}

In particular, \tapioca~performs a Monte Carlo evaluation of the integrals shown
in Equations \ref{eq:diffuse_acceptance} and \ref{eq:effective_area}. A flow
chart showing the top-level logic of the \tapioca~code, as well as the required
inputs and data sources, is shown in Figure~\ref{fig:eventloop} and described in
the remainder of this section.

 Given the location of the payload ($\phi,\lambda,h,t$), where $\phi,\lambda$ are latitude and longitude, $h$
 is the altitude above the reference Earth ellipsoid, and $t$ is the observation, and the coordinates of a
 neutrino source ($\alpha,\delta$), we calculate the \textit{geometric} area of 
 \textit{neutrino exit locations} that is potentially detectable by ANITA. This is done
 by setting a cut on the maximum view angle, $\theta_{\mathrm{view,exit}}$, at the 
 neutrino exit location on the surface as shown in
 Figure~\ref{fig:geometric_area_diagram_a}.
 
 Since the observed ANITA-IV events are extremely close to the horizon, we must accurately
 take into account the altitude of the ice and curvature of the Earth at the observed location of each event. For each
 observed ANITA-IV event, we raytrace the observed RF direction back to the continent
 and then use the BEDMAP2~\cite{bedmap2} ice dataset to calculate the altitude of the ice surface
 in the region surrounding the event in our 3D coordinate system.
 
 Using the geometric area calculated in the previous step, \tapioca~samples $N$ (the 
 number of desired Monte Carlo trials to evaluate the integral) neutrino 
 exit locations within this area consistent with a neutrino source at $(\alpha, \delta)$. 
 \tapioca~then loops over each of these sampled neutrino locations and assigns an incident neutrino
 energy given by the simulation input parameters. For each neutrino, we use the \texttt{NuTauSim} lookup-tables to sample the exit
 probability of a neutrino at this location \& neutrino energy as well as randomly sample the energy of the
 $\tau$-lepton from the \texttt{NuTauSim} distributions~\cite{2018PhRvD..97b3021A,2019PhRvD..99f3011R_Erratum}. We configured \texttt{NuTauSim}
 with the middle UHE neutrino-nucleon cross section parametrization from \cite{Connolly:2011vc} and the ALLM energy loss model~\cite{Abramowicz:1997ms} for the $\tau$ lepton (which are the defaults for this particular propagation code).
 
 We step this $\tau$ to its decay point in our full 3D coordinate system, under the assumption that
 the $\tau$ does not undergo significant energy loss in air. At the decay point, 
 we sample the decay distributions generated by PYTHIA to get the fraction of the $\tau$ energy that
 was transferred into the extensive air shower. The ZHAireS electric field model described
 in \S\ref{sec:models}, in the frequency- or time-domains (depending upon the simulation configuration),
 is then used to calculate the incident electric field at the location of the payload. The previously 
 described detector model is then applied to determine whether this trial
 passed our trigger model (i.e. $P_{\mathrm{trig}} = 1$ or $P_{\mathrm{trig}} = 0$). 
 
 The total effective area, 
 $A (t, E_\nu, \alpha, \delta, \phi, \lambda, h)$, is then calculated as
 
 \begin{equation}
   \begin{split}
     A (t, E_\nu, \alpha, \delta, \phi, \lambda, h) & \approx \\ & \frac{A_g}{N} \sum_{i=1}^{N} \hat{r}_{i,*} \cdot \hat{x}_{i,E}\,P_{i,\mathrm{exit}} P_{i,\mathrm{decay}} P_{i,\mathrm{trig}}
     \label{eq:monte_carlo_Aeff}
     \end{split}
 \end{equation}
 
 For a typical run of \tapioca~, $N \sim 10^8$ since our phase space cuts are generous
 to accurately capture the tail of the distribution.
 \texttt{tapioca} can operate in two distinct modes:
\begin{enumerate}
    \item Reconstructing the effective area associated with a specific \textit{observed} ANITA-IV event using the run, geometry, and location of the event in question.
    \item Calculating the total exposure for the entire flight using parameters sampled from throughout the flight.
\end{enumerate}
 
\begin{figure}
    \centering
    \includegraphics[width=\columnwidth]{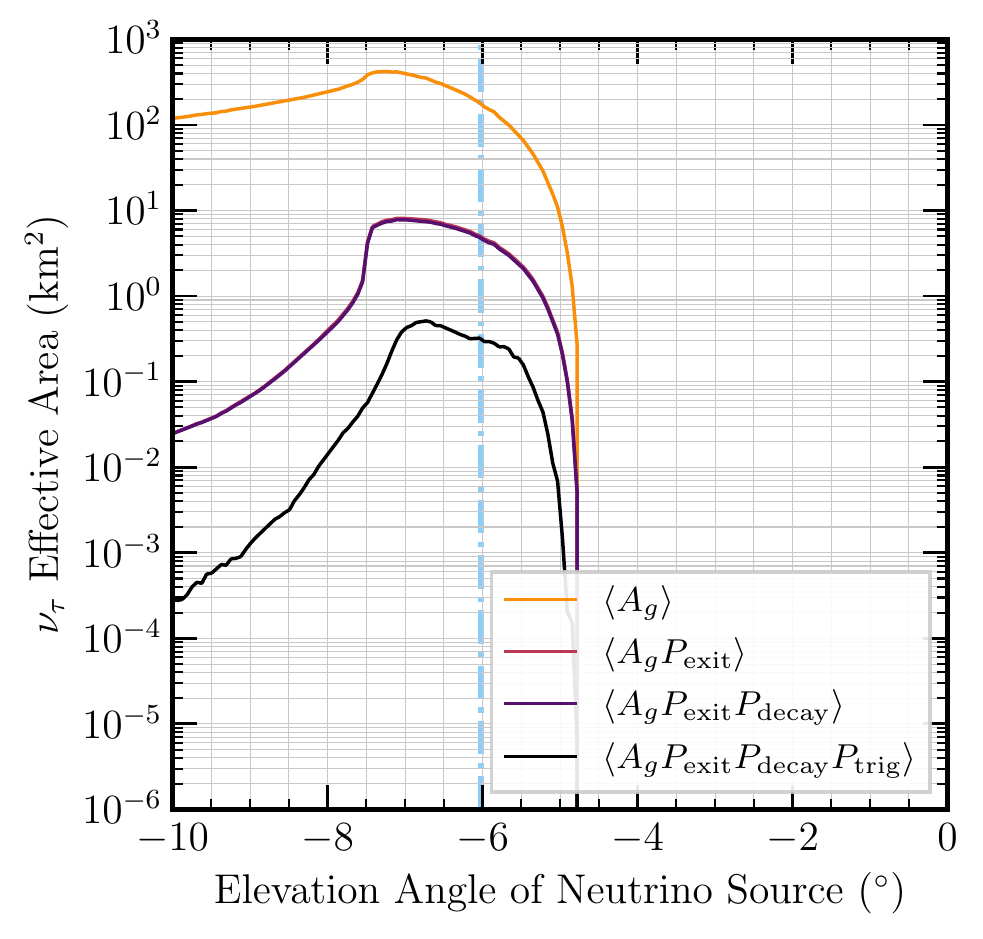}
    \caption{The effective area for $E_\nu = 10$~EeV for an average location and position in the flight
    broken down into the various components that are calculated by \tapioca~as a function
    of the elevation angle of the neutrino source below ANITA's horizontal. The dashed line indicates
    the approximate location of the geometric horizon in ANITA's coordinate system.
    \label{fig:effective_area_by_components}}
  \end{figure}

 To illustrate the various contributions to ANITA-IV's effective area, we break down
the effective area calculation for a neutrino energy of 10~EeV into the various sub-components of the Monte Carlo calculation
(Equation~\ref{eq:monte_carlo_Aeff}) in Figure~\ref{fig:effective_area_by_components}. Due to its unique
observing location at $\sim$37~km above the surface, ANITA's geometric area approaches 400~km$^2$. 
Near the horizon, the probability of an incident $\nu_\tau$ generating a $\tau$-lepton at 10~EeV is approximately
$\sim1/50$ immediately reducing the maximum effective area to $\sim$8~km$^2$; below $\sim8^\circ$, the exit probability for a $\tau$-lepton falls drastically significantly reducing the maximum potential effective area; 
typically, the $\tau$-lepton exit probability is the most relevant factor to ANITA's effective area.
Since ANITA is $\lesssim600$~km from the horizon, the probability that a $\sim10$~EeV $\tau$ decays
prior to reaching ANITA is extremely high and is not a significant contributor to the effective area as can be seen
in Figure~\ref{fig:effective_area_by_components} as the magenta and purple lines overlap significantly.

ANITA's effective area to $\nu_\tau$ sources extends above the horizon as ANITA observes the radio emission off-axis
with respect to the neutrino propagation axis. Therefore, an Earth-skimming neutrino from a source slightly above the horizon (\textit{as seen by} ANITA) can still skim the Earth, decay in the air, and be observed $\gtrsim1^\circ$ off-axis by ANITA.

ANITA is a broadband instrument and primarily designed for the detection of
Askaryan emission which has significantly more high-frequency power than an in-air cosmic ray shower. Therefore ANITA must typically observe an EAS fairly close to the Cherenkov angle in order to trigger. This was worsened by the presence of the notch filters (described in \S \ref{sec:detector_model}) that preferentially removed low-frequencies (200-500~MHz) where the spectrum of an EAS has maximum spectral power density.
This \textit{on-cone} geometric factor in the detection model shows up in the trigger probability calculation
and further reduces the effective area calculation by an order of magnitude.  

\section{Results}
\label{sec:results}


\subsection{Interpreting ANITA events as upgoing tau neutrinos}
\label{sec:anita_events_as_taus}

In \S\ref{sec:neutrino_source}, we consider the observational evidence that
the four near horizon events originated from upgoing $\tau$-neutrinos via the $\tau$-induced
EAS channel. This includes
considering the implied neutrino source directions, energies, and a spectral analysis against
simulations. We find that the events are not inconsistent with the tau neutrino hypothesis.
In \S\ref{sec:diffuse} and \S\ref{sec:point_source}, we further consider the implications of the tau neutrino hypothesis
by considering both an isotropic, diffuse flux of tau neutrinos and point sources -- both steady and transient -- of tau neutrinos. 
This includes comparisons to other experimental searches for ultrahigh-energy tau neutrinos.

\subsubsection{Neutrino Source Parameters}
\label{sec:neutrino_source}

Uncertainties in the location of the neutrino source on the sky with right-ascension ($\alpha$) and declination $(\delta)$ are dominated by the cone-shaped beam of the upgoing air shower radio emission, which itself varies with the $\tau$-lepton decay altitude and the zenith angle of the shower. This motivates the use of a forward modeling approach to reconstruct the location of the neutrino source and the corresponding energy of the neutrino that is most consistent with our observed events.

We developed a
Markov Chain Monte Carlo simulation to reconstruct the posterior distributions of $(E_\nu, \alpha, \delta)$;
in particular, we use the \texttt{emcee} package~\cite{2013PASP..125..306F} that implements an affine-invariant ensemble MCMC sampler~\cite{2010CAMCS...5...65G}.
For this reconstruction, we use the following likelihood function, implemented in \tapioca,~that captures the entire process of neutrino emission
to the observed radio-frequency signal:

\begin{equation}
    \mathcal{L}(E_\nu, \alpha, \delta, t) \quad \propto \quad 
    A_g\,P_{\mathrm{exit}}\,P_{\mathrm{decay}}\,\mathcal{L}_{\theta}\,\mathcal{L}_{\phi}\,\mathcal{L}_{\mathrm{waveform}}
    \label{eq:source_likelihood}
\end{equation}
where:
\begin{conditions*}
  A_g & the geometric area for a given $(\alpha, \delta, t_{\mathrm{event}}, \vec{x}_{\mathrm{payload}})$, \\
  P_{\mathrm{exit}} & the probability that this neutrino generates a $\tau$-lepton that leaves the Earth, \\
  P_{\mathrm{decay}} & the probability that the $\tau$-lepton decayed before ANITA, \\
  \mathcal{L}_{\theta} & a Gaussian likelihood for the observed RF elevation angle, \\
  \mathcal{L}_{\phi} & a Gaussian likelihood for the observed RF azimuth angle, \\
  \mathcal{L}_{\mathrm{waveform}} & a sample-by-sample Gaussian likelihood for the residuals between the simulated waveform and the observed waveform.
\end{conditions*}

Since the neutrino flux distribution at these energies is unobserved, we repeat this likelihood optimization for different priors on the neutrino spectrum. We assume a 
generic power law neutrino flux distribution, $E_{\nu}^{\gamma}$, between 0.1~EeV and 1000~EeV and reconstruct the neutrino parameters for discrete values $\gamma \in \left\{-3,-2,-1\right\}$ to accommodate a range of cosmogenic 
and astrophysical neutrino models.

\begin{figure*}
  \centering
  \includegraphics[width=1.5\columnwidth]{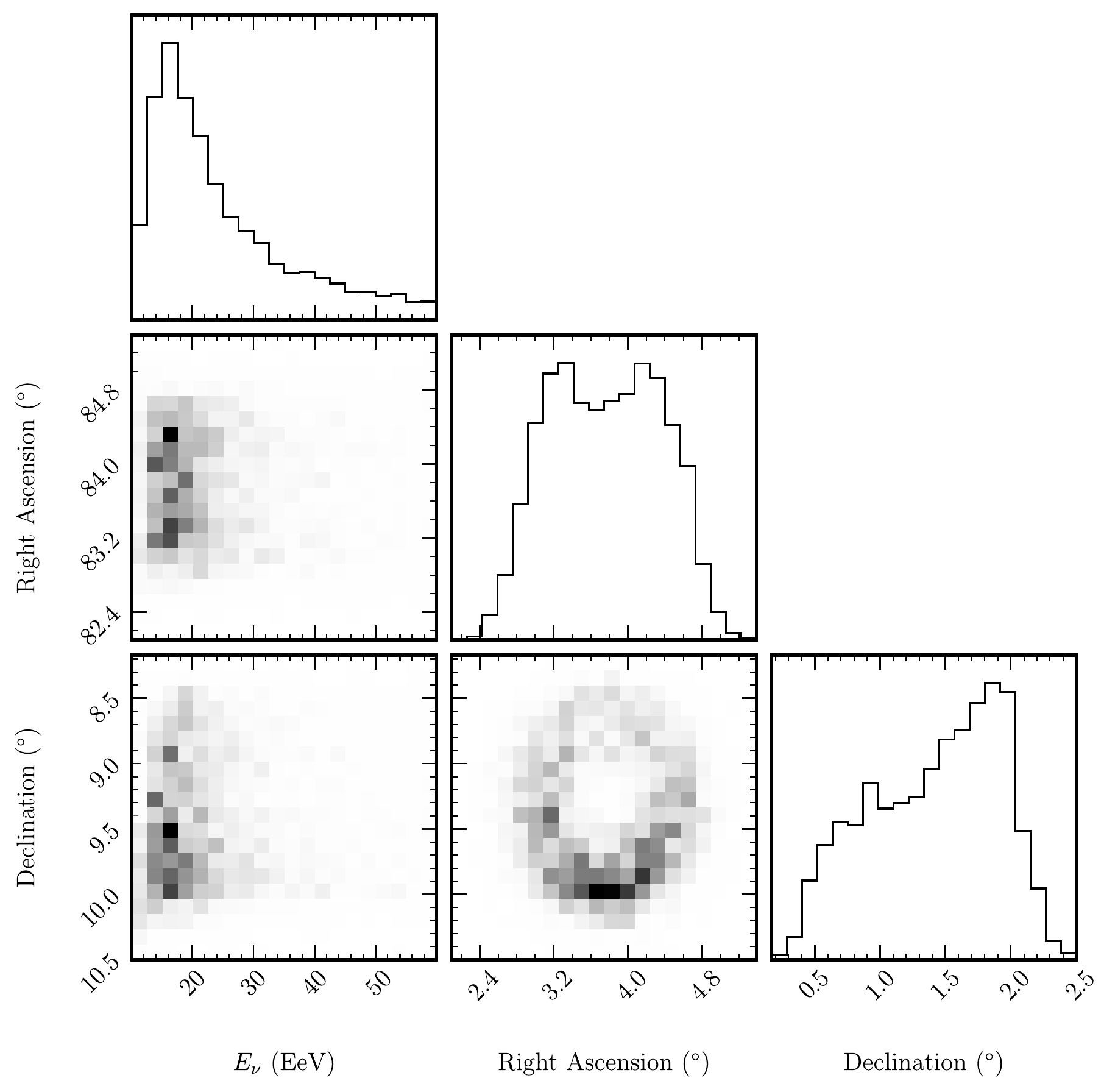}
  \caption{The posterior distributions of the neutrino energy and neutrino source locations, $(E_\nu, \alpha, \delta)$, 
  for Event 72164985 under a $\gamma=-2$ hypothesis as reconstructed by the \texttt{emcee}~\cite{emcee} Markov Chain Monte Carlo using the likelihood function 
  in Equation~\ref{eq:source_likelihood}. We note that the right-ascension and declination have been shifted by a random (constant) offset for this publication. The true reconstructed sky coordinates will be published in a follow-up
  paper by the ANITA collaboration.}
  \label{fig:Enu_posteriors_721}
\end{figure*}

We use broad ($6^{\circ}$ wide) uniform priors on $\alpha$ and $\delta$ centered around the sky location of the observed RF of each event. As an example, we consider the posterior distributions of $(E_\nu, \alpha, \delta)$ for Event 72164985 (Figure~\ref{fig:Enu_posteriors_721}). The most likely neutrino energy depends strongly on the assumed neutrino spectral index, $\gamma$. A harder spectra ($\gamma \sim -1$) is more likely to generate higher energy neutrinos that have more phase space for producing a decaying $\tau$ with sufficient shower energy to trigger ANITA; equivalently, a softer spectrum ($\gamma \sim -3$) strongly disfavors the production of higher energy $\nu_\tau$'s and therefore requires an upward fluctuation in the $\tau$ energy and shower fraction in order to be detected by ANITA. \textbf{}For $\gamma=-2$, the 50\% quantile in reconstructed neutrino energy for Event 72164985 is $15.1$~EeV with lower and upper 1-$\sigma$ quantiles of $7.6$~EeV and $42.4$~EeV, respectively.

We note that the energies given in Table I of~\cite{2021PhRvL.126g1103G} are generic cosmic-ray shower energies (not neutrino energies) that were estimated using standard cosmic-ray shower models as opposed to the dedicated upward $\tau$ EAS simulations used in this work.

The reconstructed neutrino parameters for all of the four events are shown in Table~\ref{tab:neutrino_parameters} under the various assumptions for $\gamma$.
This MCMC, which forward models the entire process from incident neutrino to detection by ANITA, includes uncertainties in the detector models as described in \S\ref{sec:models}, as well as the uncertainty in the observed event parameters.


\renewcommand{\arraystretch}{1.5}
\begin{table*}
  \caption{\label{tab:neutrino_parameters} The most-likely reconstructed neutrino energies, using the MCMC approach described
  in \S~\ref{sec:neutrino_source}, for various priors on the neutrino flux.}
\begin{ruledtabular}
\begin{tabular}{c | cccc}
Event  & $E_{\nu,\gamma=-1}$ (EeV)  & $E_{\nu,\gamma=-2}$ (EeV)  & $E_{\nu,\gamma=-3}$ (EeV) \\
  \hline
  4098827  & $49.8^{ +80.3 }_{ -37.7 }$ & $12.5^{ +29.9 }_{ -7.4 }$ & $5.2^{ +6.0 }_{ -2.5 }$ \\
  19848917  & $31.9^{ +76.0 }_{ -24.5 }$ & $5.2^{ +11.0 }_{ -2.9 }$ & $2.6^{ +3.1 }_{ -1.1 }$ \\
  50549772  & $45.4^{ +83.4 }_{ -34.4 }$ & $8.8^{ +19.5 }_{ -4.9 }$ & $4.3^{ +4.8 }_{ -2.1 }$ \\
  72164985  & $60.3^{ +88.9 }_{ -38.2 }$ & $15.1^{ +27.3 }_{ -7.6 }$ & $8.9^{ +10.5 }_{ -4.5 }$ \\
\end{tabular}
\renewcommand{\arraystretch}{1}
\end{ruledtabular}
\end{table*}



Since ANITA orbits the South Pole, ANITA's elevation (azimuth) resolution is almost purely converted into declination (right-ascension) resolution. Figure~\ref{fig:Enu_posteriors_721} shows that
the most likely neutrino source location $\alpha$ and $\delta$ is distributed as an annulus on the sky with an angular radius of $\sim1^\circ$. This is due to the conically shaped radio emission from the EAS
which has a Cherenkov angle of approximately $1^\circ$ in air. 

Since ANITA only observes the radio emission at one \textit{point} on the
Cherenkov cone, the shower energy and the azimuthal angle around the shower axis
(equivalently, \textit{which} side of the cone you are observing) are degenerate
as higher-energy showers may be observed with less radio power if the Askaryan
and geomagnetic components are anti-aligned. Equivalently, if the Askaryan and
geomagnetic components are aligned, this will result in an increase in the
observed RF emission. It may be possible for ANITA to break this degeneracy
through a detailed analysis of event polarization and the local geomagnetic
field, but this is challenging to simulate due to the geometry of the near
horizon events (see \S \ref{sec:simulation_challenges}). The results of the MCMC
shown in Figure~\ref{fig:Enu_posteriors_721} and
Table~\ref{tab:neutrino_parameters} indicate that ANITA has approximately
$\pm 1^\circ$ resolution in right-ascension and $\pm 0.5^\circ$ resolution in
declination for the $\tau$-induced extensive air shower channel. We note that
this is not ANITA's resolution for reconstructing the observed direction of the
RF emission (which is typically $\sim0.2^\circ$,
see~\cite{2021PhRvL.126g1103G}); the additional uncertainty is due to the
uncertainty for the azimuthal angle around the shower axis of the EAS as well as
the reconstruction of the off-axis angle of observation.

\subsubsection{Spectral Analysis}

In this section, we compare the observed spectra against simulated spectra for upgoing $\tau$-induced
EAS produced using ZHAireS~\cite{2012APh....35..325A}.
The spectral shape of an upgoing air shower radio pulse can be approximated as a falling exponential above $\sim300$~MHz~\cite{2010PhRvL.105o1101H,Schoorlemmer:2015afa, 2013AIPC.1535..143A}. The amplitude and exponential constant are dependent on the shower energy, the off-axis angle, the decay altitude of the $\tau$-lepton, and the zenith angle of the shower. The decay altitude and the zenith angle changes the atmospheric profile
 over which the shower develops, which affects the coherence of the radio emission,
 resulting in a different off-axis dependence as well as a lower required energy since the shower can develop closer to ANITA depending upon the decay length of the $\tau$ lepton.

\begin{figure}
  \centering
  \includegraphics[width=1\columnwidth]{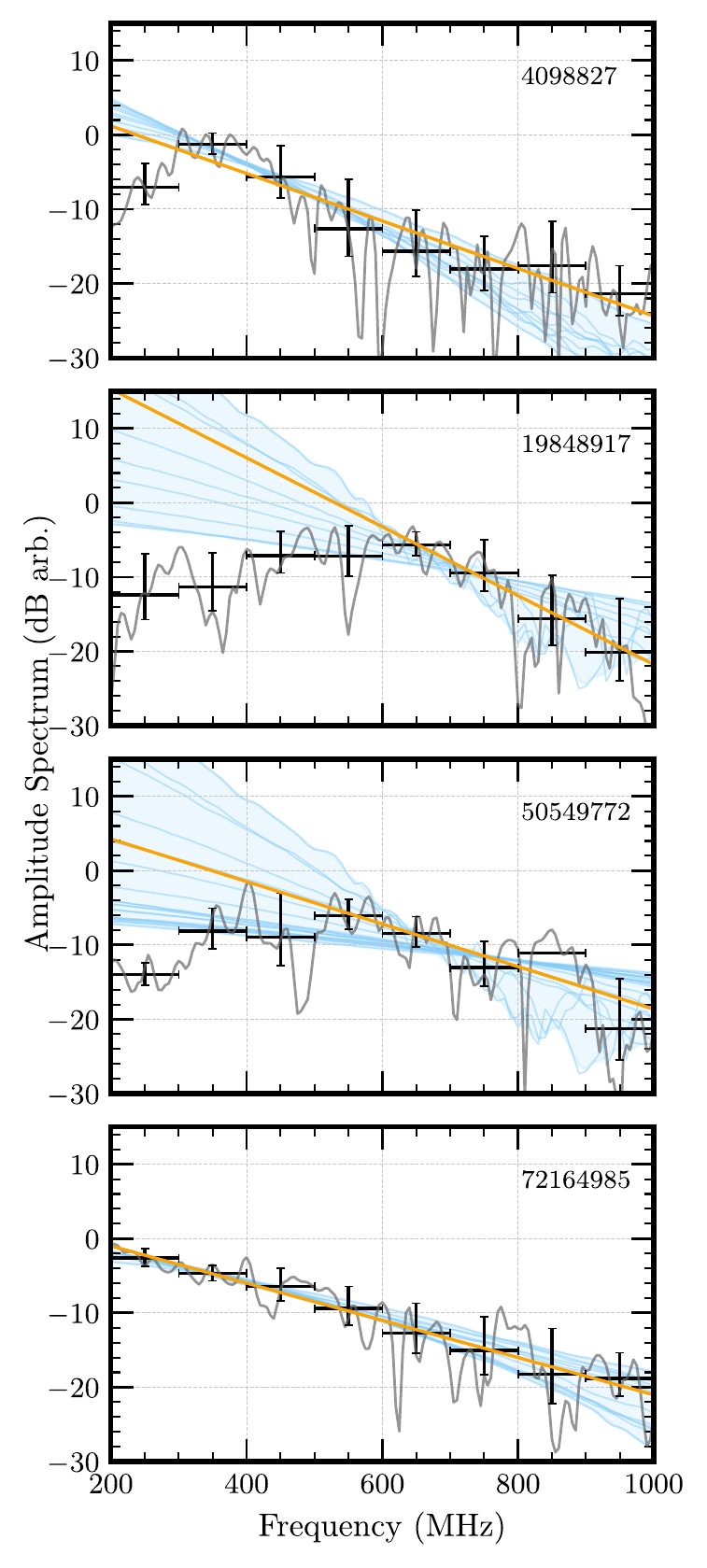}
  \caption{The deconvolved electric field amplitude spectrum for each event
  produced using the CLEAN deconvolution algorithm described in \cite{2021PhRvL.126g1103G}.
  The amplitude spectrum (gray) is shown along with the average over consecutive independent 100~MHz bins (black).
  Each set of averages (black) was fit with an exponential form (orange), $A\exp(\gamma(f - 300))$. For 4098827 and 72164985, the fit was performed from 300~MHz up to 1000~MHz. For 19848917 and 50549772, the fit was performed
  over the frequency region above the peak ``turnover'' (typically 500-600~MHz).
  Several different ZHAireS simulated upgoing $\tau$ spectra are also shown (light blue) for a variety of decay altitudes and
  zenith angles that could be consistent with these events.
  \label{fig:deconvolved_waveforms}}
\end{figure}

Figure~\ref{fig:deconvolved_waveforms} shows the deconvolved electric field spectra for each event~\cite{2021PhRvL.126g1103G}.
Since the  spectrum of extensive air showers is well approximated by an 
exponential spectrum above 300~MHz, we also fit each deconvolved spectrum with an
exponential function, $A(f) = A_{\mathrm{300}}\,e^{\gamma(f - 300)}$, with $f$ in MHz where 
$\gamma$ is the spectral index. We also show a range of electric field spectra simulated by ZHAireS 
to demonstrate the range of spectral indices observed in upgoing $\tau$-induced air showers (light blue).

Event~72164985, which is closest to the
horizon, is an excellent match to simulated upgoing EAS signals as well as the expected exponential spectral profile. Events 19848917 and 50549772 show a clear reduction in spectral power at 
low frequencies ($\lesssim500~\mathrm{MHz}$). It is possible that atmospheric effects due to the long distance propagation from close to
the horizon could create anomalous spectra like is seen in these events. Possible physical explanations for this missing low-frequency power are discussed in \S\ref{sec:low_frequency_anomaly}. For the purpose of our simulations, 
we do not include or model this low-frequency reduction and only fit the frequency range above the location of max spectral power (typically $300-500~\mathrm{MHz}$) for Events 19848917 and 50549772. Above $\sim500$~MHz, the events appear to agree with our simulations as well as with the expected exponential shape.

\begin{figure}[htbp]
  \centering
  \includegraphics[width=1\columnwidth]{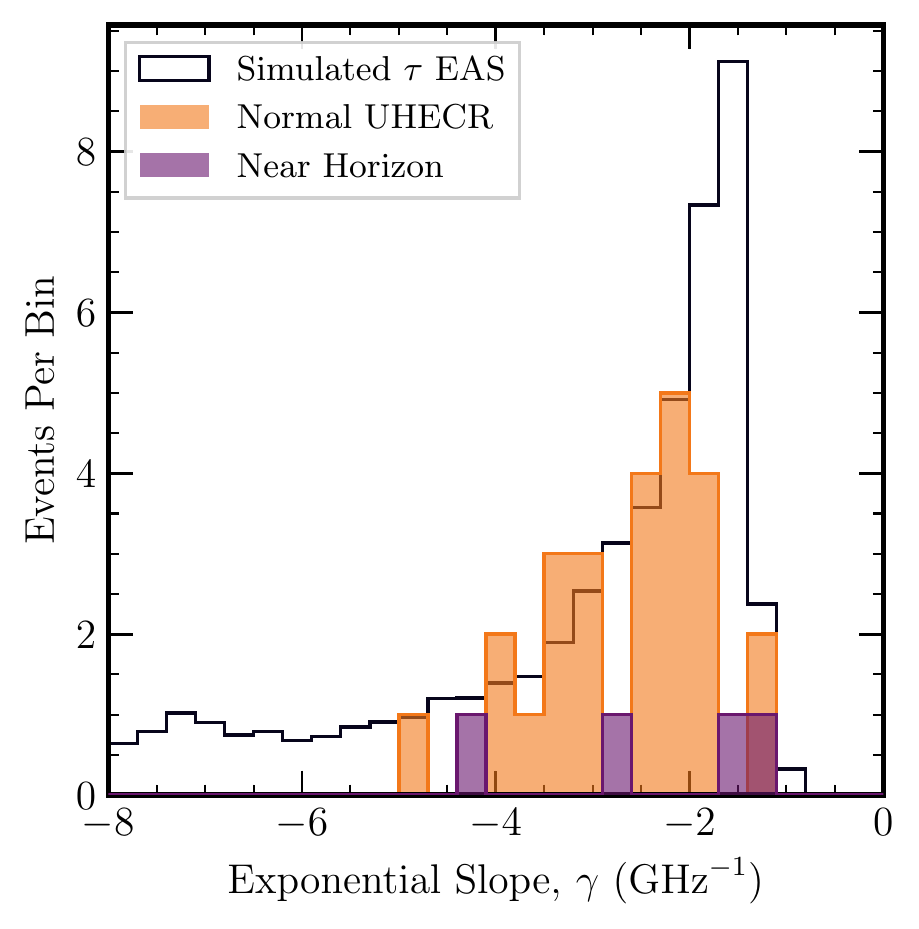}
  \caption{The exponential radio-frequency spectral slope distribution of the waveforms 
  of the regular (reflected) cosmic
  rays (orange) observed by ANITA-IV compared against the spectral indices of the four near-horizon
  $\tau$-like events (purple) compared to an arbitrary scaled distribution of $\tau$ events simulated with \tapioca.
  We note that to be consistent with previous ANITA publications, we have reused $\gamma$ here as the spectral slope (with
  units of inverse frequency) of the radio-frequency waveform whereas early (to be consistent with other published work) we also used $\gamma$ as the exponent
  in the power law neutrino flux distribution.}
  \label{fig:spectral_index_distribution}
\end{figure}

We use \tapioca~to generate a sample of simulated $\tau$-EAS events that passed ANITA's
trigger so that we can compare these four events against the expected broader population of $\tau$ EAS.
The spectral indices of each of the four
near horizon events compared to the full population of normal (reflected and stratospheric) observed cosmic rays, as well as the
simulated events, is shown
in Figure~\ref{fig:spectral_index_distribution}. We do not necessarily expect that the spectral index of reflected UHECRs should match that of the simulated $\tau$-induced EASs due to the
  different event geometries as well effects from the reflection of the radio emission off the ice.

We perform a Kolmogorov-Smirnov (KS) test to determine whether the spectral indices of
the four near horizon events 
are consistent with the underlying population of \textit{normal} UHECRs or the simulated $\tau$ distributions. The results 
of this test are shown in Table~\ref{table:ad_test_spectra}.

\begin{table}[h]
  \caption{\label{table:ad_test_spectra} The results of a Kolmogorov-Smirnov test on the spectral index, 
  $\gamma$, of the near horizon events compared against the sample of regular UHECRs and simulated $\tau$
  EAS.}
\begin{ruledtabular}
\begin{tabular}{cl}
Test &  p-value \\
  \hline
  Near horizon against regular UHECRs  & 0.48 \\
  Near horizon against simulated $\tau$ EAS (ZHaireS) & 0.45\\
\end{tabular}
\end{ruledtabular}
\end{table}

In both cases, the KS test finds no evidence that the four near horizon events are not drawn from the underlying distribution. The p-value of both cases is approximately 0.5 which is an order of magnitude worse than would be needed to reject the hypothesis
that the four NH events came from each underlying distribution at a 5\% significance. 

Considering both the implied neutrino parameters $(E_\nu, \alpha, \delta)$ and the spectral evidence, we conclude that these events are not inconsistent
with $\tau$-induced EAS.

\subsection{Diffuse Flux Limits}
\label{sec:diffuse}

\begin{figure}
  \includegraphics[width=1\columnwidth]{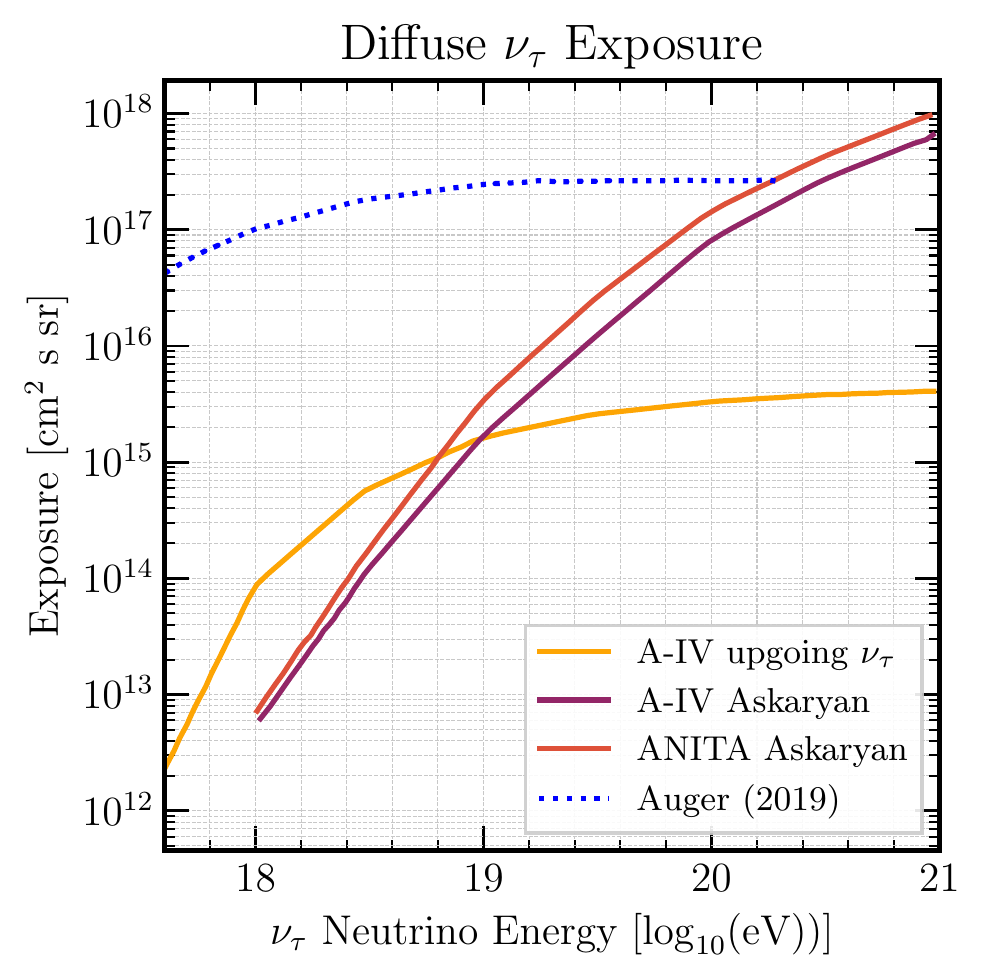}
  \caption{The (single-flavor) exposure to $\nu_\tau$ for ANITA's EAS channel as well as 
its Askaryan in-ice detection channel~\cite{2019PhRvD..99l2001G} compared against $\nu_\tau$ exposures from the Pierre Auger Observatory~\cite{2017ICRC...35..972Z}.}
  \label{fig:a4_diffuse_exposure}
\end{figure}

Figure~\ref{fig:a4_diffuse_exposure} shows the exposure of ANITA-IV to a diffuse $\nu_\tau$ flux via both the Askaryan~\cite{2019PhRvD..99l2001G} and upgoing EAS channels (this work). The upgoing EAS channel dominates ANITA's $\nu_\tau$ exposure at energies below $\sim10^{19}$~eV above which the Askaryan channel dominates. Due to the short flight of ANITA-IV ($\sim28$~days), neither the Askaryan or EAS $\nu_\tau$ exposure is comparable with IceCube or Auger at energies below $\sim10^{19.5}$~eV (the combined limit of ANITA I-IV sets a
stronger limit than is shown in Figure~\ref{fig:a4_diffuse_exposure}). The significant
discrepancy in the total exposure rules out a diffuse isotropic $\nu_\tau$ flux origin for the four ANITA-IV near horizon events under the Standard Model. This is the same conclusion reached for the two \textit{steeply upgoing} events observed in ANITA-I and ANITA-III~\cite{2019PhRvD..99f3011R,2019PhRvD..99f3011R_Erratum} and consistent with the expected sensitivity of the three flights~\cite{Wissel:2019ot}.

\subsection{Point Sources}
\label{sec:point_source}

In this section, we present ANITA-IV's sensitivity to astrophysical 
neutrino point sources, including transient sources that may have only been active for
extremely short durations. Due to its unique viewing location in the stratosphere, ANITA typically
has a large instantaneous effective area that partially compensates for its $\sim30$~day observing time.

\begin{figure}
    \centering
    \includegraphics[width=\columnwidth]{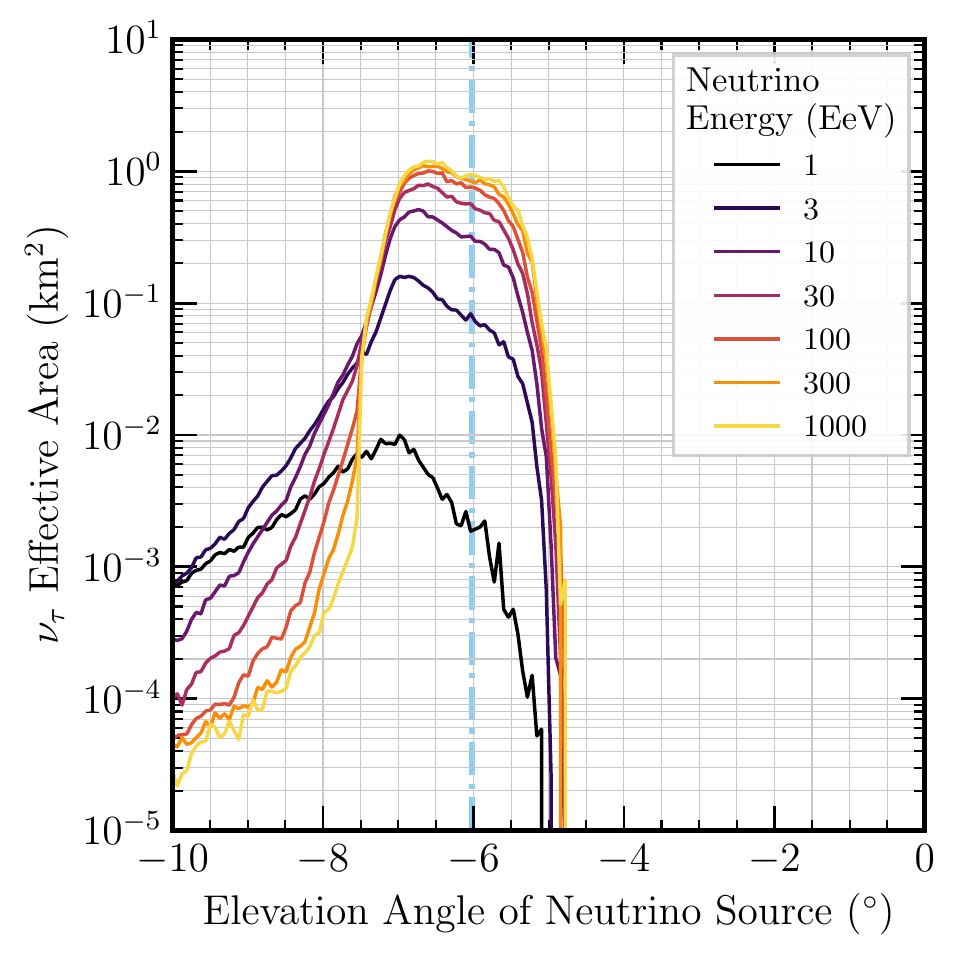}
    \caption{The effective area vs. neutrino source elevation angle for various
      tau neutrino energies from 1~EeV to 1000~EeV. The
      dashed blue line shows the approximate location of the geometric horizon (averaged over the flight)\footnote{the actual location of the observed radio horizon
     differs from the geometric horizon due to refraction of the radio emission during $\sim600$~km of propagation and depends on the payload's altitude, ice thickness
     near the horizon, and the instantaneous atmospheric conditions.}.
      }
    \label{fig:effective_area_by_energy}
\end{figure}

The effective area as a function of the elevation angle of the neutrino source on the sky for a range of 
energies from 1~EeV to 1000~EeV is shown in Figure~\ref{fig:effective_area_by_energy}. ANITA's 
EAS sensitivity to $\nu_\tau$ turns on at several hundred PeV with a peak effective area of
$\mathcal{O}$(1~km$^2$) at the highest energies. By $\sim$300~EeV, ANITA's effective area begins to saturate as the trigger 
probability tends to $\sim1$ for events at these energies. ANITA's peak sensitivity to neutrino sources occurs
when the neutrino source is roughly $\sim1^\circ$ below the horizon since this allows for a larger
geometric area of possible neutrino exit locations for neutrinos from that source to be geometrically visible by ANITA and observed close to the Cherenkov angle (Figure~\ref{fig:effective_area_by_components} and Figure~\ref{fig:effective_area_by_energy}).

\begin{figure}
    \centering
    \includegraphics[width=\columnwidth]{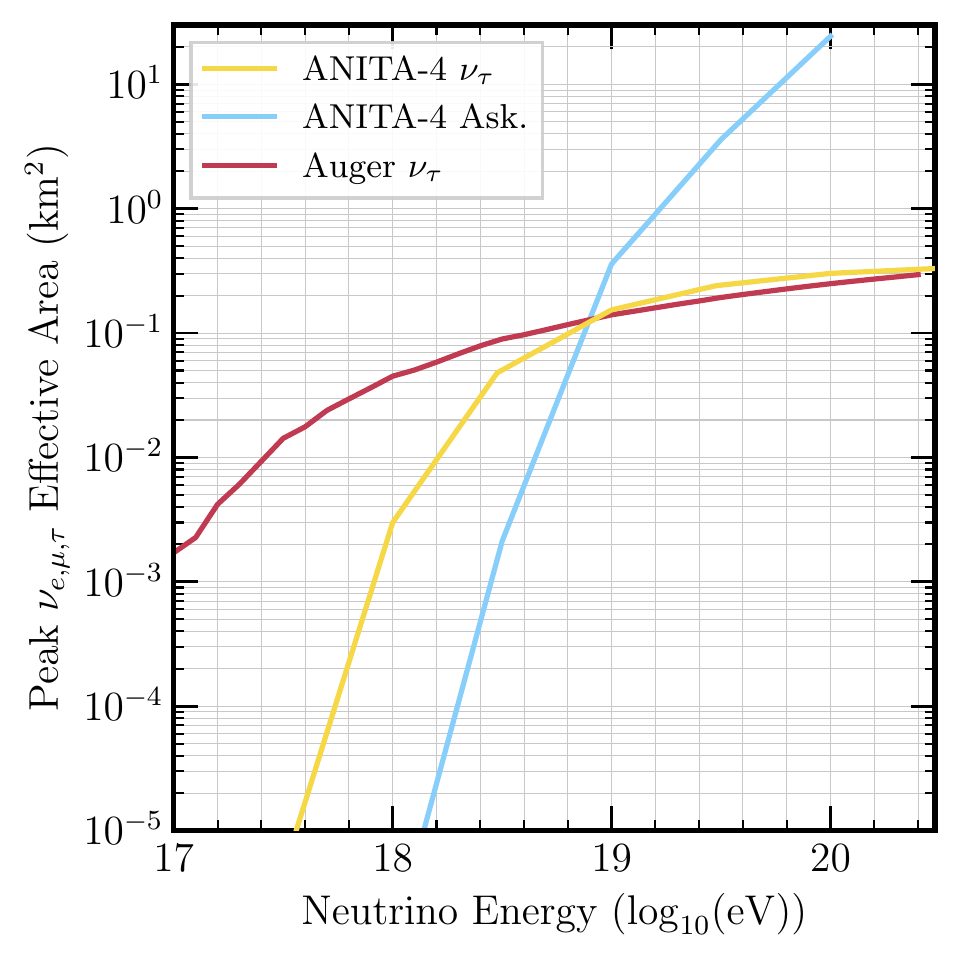}
    \caption{The peak \textit{all-flavor} (1:1:1) effective area (over elevation angle) as a function of neutrino energy for the ANITA-IV air shower channel, the ANITA-IV Askaryan channel, and the Pierre Auger Observatory's upgoing $\nu_\tau$ channel. The Auger curve was extracted using published data in~\cite{2019JCAP...11..004A}.}
    \label{fig:peak_effective_area}
\end{figure}

The \textit{peak} effective area of ANITA-IV as a function of incident neutrino
energy for both the Askaryan channel (single-flavor effective area assuming a 1:1:1 flavor ratio)
and the $\tau$ air shower
channel is shown in Fig.~\ref{fig:peak_effective_area}. The effective area in the $\nu_\tau$ EAS channel exceeds that of the Askaryan channel, below $10^{19}$~eV and significantly lowers ANITA's threshold energy for $\nu_\tau$ detection down to $\sim0.3$~EeV. The sensitivity of the Askaryan channel exceeds the $\nu_\tau$ EAS channel above $10^{19}$~eV and can approach $\sim30$~km$^2$ at the highest energies. We also show the Pierre Auger Observatory's
upgoing $\nu_\tau$ effective area over the same energy range using the published data from~\cite{2019JCAP...11..004A}. 

\subsubsection{Consistency with $\nu_\tau$}

We compare where our four observed events occur within the expected distributions of the elevation angles
of $\tau$-induced EAS events as simulated by \texttt{tapioca}. Using the
parameters (payload location, time, ice thickness, etc.) at the time of each observed near horizon event, we generate a large random sample of simulated $\nu_\tau$ air shower events that would have been detected by ANITA by sampling
the calculated effective area shown in Figure~\ref{fig:effective_area_by_energy}. We use the energy curve closest
in energy to the reconstructed neutrino energies shown in Table~\ref{tab:neutrino_parameters} assuming a $E^{-2}$ prior
on the neutrino flux energy spectrum.

Given this distribution of elevation angles, we perform a series of Kolmogorov-Smirnov tests to determine if the observed events are consistent with the simulated event distributions; the p-values
for the KS tests are shown in Table~\ref{table:ad_test_location}. 


  

\begin{table}[h]
     \caption{The p-value of a Kolmogorv-Smirnov test for rejecting the hypothesis that these events are drawn from the simulated distribution
     of events from \tapioca.}
\begin{tabular}{l | c}
    \label{table:ad_test_location}
     Event & KS p-value \\ \hline
     4098872 & 0.95 \\
     19848917 & 0.60 \\
     50594772 & 0.72 \\
     72164985 & 0.85 \\ \hline 
     All Events & 0.19

\end{tabular}
\end{table}

For each observed event, there
is no evidence to reject the hypothesis that the observed events are taken from the simulated distribution of elevation angles of $\tau$-induced EAS events. Under the model presented in this work, all four events taken together are observed at elevation angles that are \textit{not inconsistent} with $\tau$-induced EAS.

\subsubsection{Field of View}
ANITA has a relatively narrow 
instantaneous field-of-view on the sky (a $\sim1^\circ$ wide band in elevation angle) for
which it has a large effective area, so the effective area over time for a given sky coordinate is not constant
due to the orbital movement of the payload and the sidereal motion of the source on the sky. The ANITA payload typically completes several full orbits around the geographic south pole (i.e. $\phi \in [-180, +180]$) with a latitude that varies between $-90^{\circ}$ and $\lesssim-75^{\circ}$

\begin{figure*}
    \centering
   \includegraphics{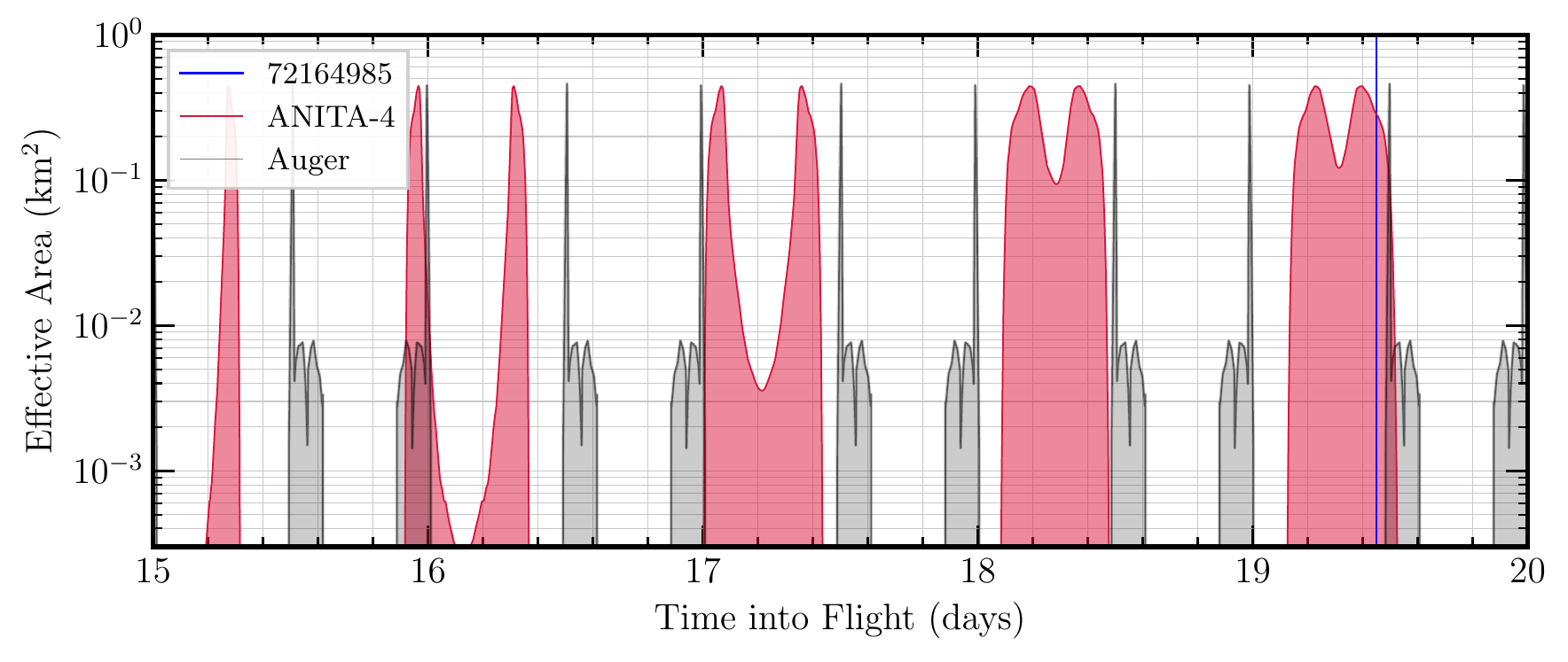}
    \caption{The time evolution of the effective area of ANITA-IV $\nu_\tau$ (pink) and Auger (grey)~\cite{2019JCAP...11..004A} in the direction of the peak source location corresponding to event 72164985 (see Table~\ref{tab:neutrino_parameters}). Event 72164985 occurred at the time indicated by the blue line at day $\sim19.5$.
    The Auger curves (simulated by us) were performed using the published curves in~\cite{2019JCAP...11..004A}.}
    \label{fig:721_time_curve}
  \end{figure*}

The instantaneous effective area at 1~EeV for a 5-day period encompassing the detection
of Event~72164985 is shown in Figure~\ref{fig:721_time_curve} along with the effective area
of the Pierre Auger Observatory~\cite{2019JCAP...11..004A}. As ANITA is also \textit{orbiting}
the continent, the shape and duration of each daily viewing period changes on a day-to-day basis.
The time at which ANITA observed event~72164985
is shown with a blue vertical line. All four near horizon events were observed by ANITA during a window
when they \textit{were not} visible by Auger and occurred close to the daily peak in $\nu_\tau$ effective area.

\begin{figure}
   \includegraphics[width=1\columnwidth]{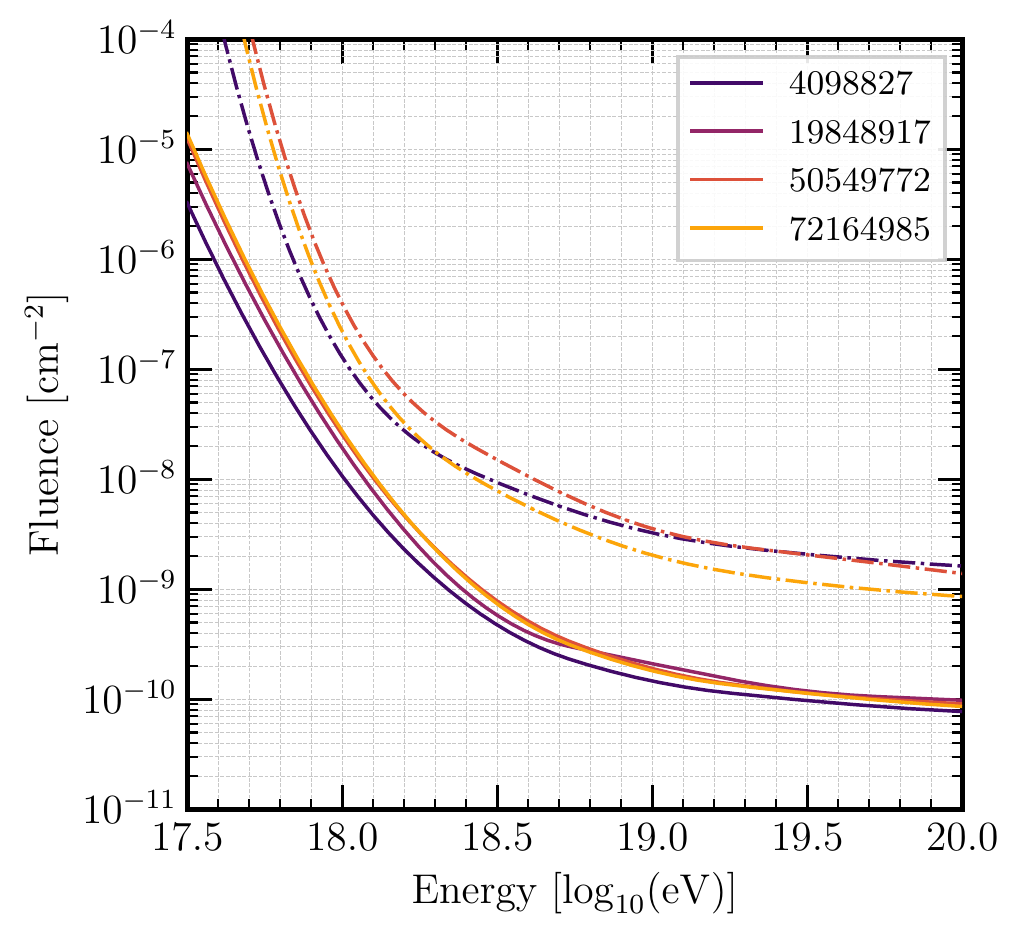}
   \caption{ANITA's sensitivity to short-duration ($<15$~minute) (solid)
   and long duration ($>1$ day) (dashed) transient neutrino sources at the
   location of each of the four near-horizon events observed in ANITA-IV.
   \label{fig:a4_ps_sensitivity}}
 \end{figure}

As shown in Figure~\ref{fig:721_time_curve}, ANITA's effective area to a given neutrino source location on the sky can be large, 
but varies significantly as a function of time since the visible portion of the sky changes and ANITA's effective area depends
strongly on elevation angle.
Therefore, ANITA can set different sensitivity limits on the point source flux depending upon the
duration of the transient source. The \textit{instantaneous} single event sensitivity (SES) limit set by 
ANITA-IV for short-duration ($<15$~minute) and long duration ($>1$~day) transient
neutrino sources occurring at the location of the four observed near-horizon events is shown in Figure~\ref{fig:a4_ps_sensitivity} (this SES limit is estimated using the method in~\cite{2021JCAP...04..017A}). Since each event was
observed at a different location on the $A_{\mathrm{eff}}(\delta)$ curve, and since each of
the sources moves in and out of ANITA's field of view with different transit rates, the
strength of the SES is different for each neutrino candidate location and for different event durations. 
However, since each event was observed very close to the peak in ANITA's time-varying $\nu_\tau$ effective area, the
short-duration SES limits for each event are similar

\begin{figure}
   \includegraphics[width=1\columnwidth]{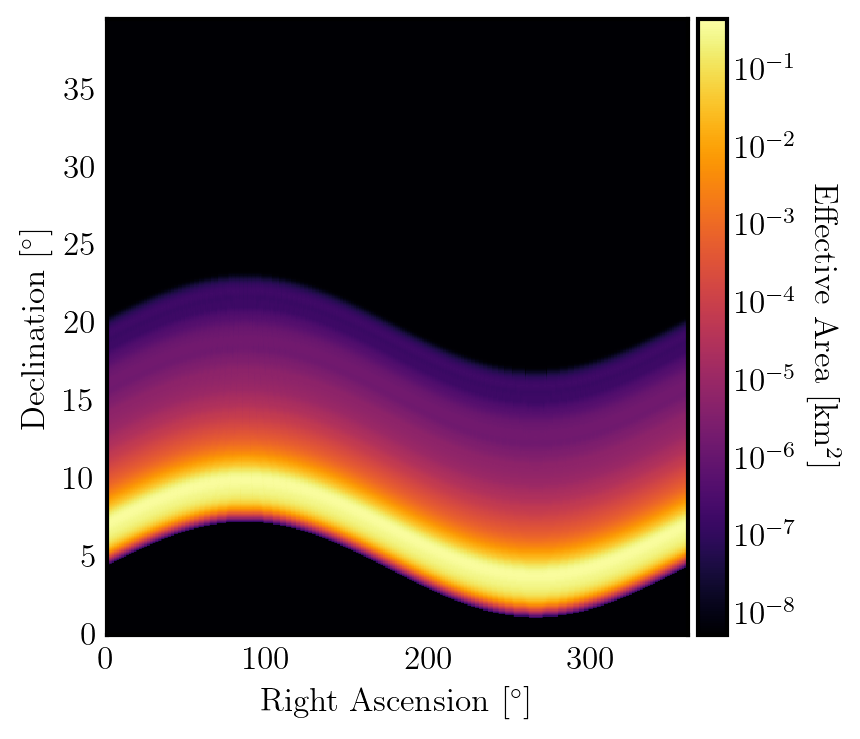}
   \caption{An instantaneous sky map of ANITA-IV's $\nu_\tau$ effective area
   over right-ascension and declination at the time of observation of Event~72164985 for a neutrino energy of $100$~EeV.}
   \label{fig:skymap}
 \end{figure}


\subsubsection{Comparison with other Observation Channels}
  
Under the assumption that ANITA-IV observed 3-4~$\nu_\tau$ events (Figure~\ref{fig:true_number_of_events}) from a population
of transient neutrino sources, we calculate the number of events that should have been
observed by the Pierre Auger Observatory (Auger) as well as ANITA-IV's Askaryan neutrino channel (which
observed 1 candidate neutrino event consistent with background)~\cite{2019PhRvD..99l2001G}. 

\begin{figure}
  \includegraphics[width=1\columnwidth]{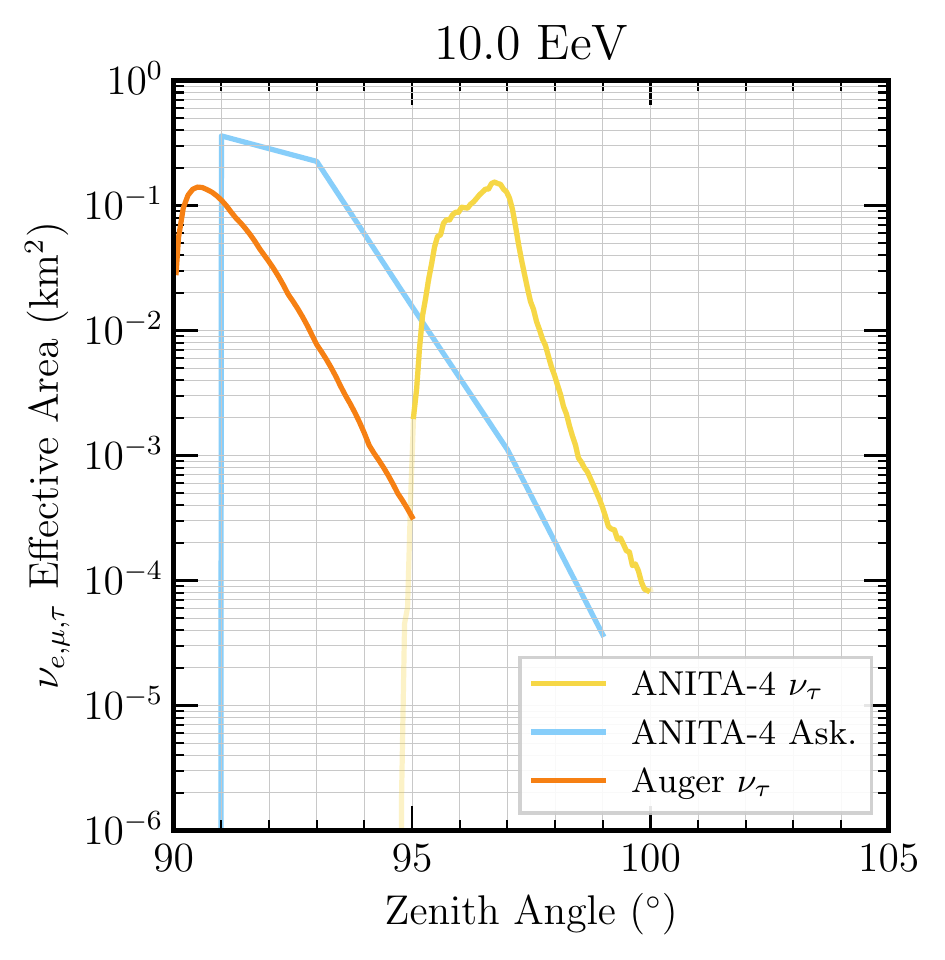}
  \caption{Single flavor effective area for the ANITA air shower channel (this work) and Askaryan channel~\cite{2021JCAP...04..017A} as well as the Auger Earth-skimming $\nu_\tau$ channel. The Auger
  curve was produced using published data from~\cite{2019JCAP...11..004A}.}
  \label{fig:field_of_view}
\end{figure}

The field of view (FoV) of ANITA's $\nu_\tau$ EAS channel is compared to the ANITA Askaryan channel and Auger's $\nu_\tau$ channel in Figure~\ref{fig:field_of_view} and Figure~\ref{fig:skymap}.
The FoV of both Earth-skimming $\nu_\tau$ channels (ANITA and Auger) are both $\sim5^\circ$, as it is strongly driven
by the exit probability of a $\tau$-lepton from a $\nu_\tau$ which is mostly independent of each detector. The peak effective area at 10~EeV is comparable for all three channels, with the Askaryan channel a factor of two higher than the two Earth-skimming ${\nu_\tau}$ channels.

We compare ANITA and Auger's sensitivity to a population of transient neutrino sources using a flux-model independent approach. 
We calculate the sensitivity of ANITA's $\nu_\tau$ and Askaryan channels, as well as those of Auger, in logarithmic
energy bins between 0.1~EeV and 1000~EeV. We then compare the fluence sensitivity for transients of various
durations from 1~second to half-day timescales, as well as for different full-sky transient rates
varying from 1~per-month to several thousand per day. While not an exhaustive search of the parameter space, 
this covers a representative sample of short- and long-duration transients that are potentially detectable by ANITA without detection
by Auger. We simulate the period between May 1st, 2008 to August 31st, 2018 which corresponds to the 
published exposure and effective curves in~\cite{2019JCAP...11..004A}. This corresponds to a total of 
exposure time $T_{\mathrm{auger}} \sim 3700$ days during which ANITA-IV flew for 28 days starting in
December 2016.

\begin{figure}
    \centering
    \includegraphics[width=\columnwidth]{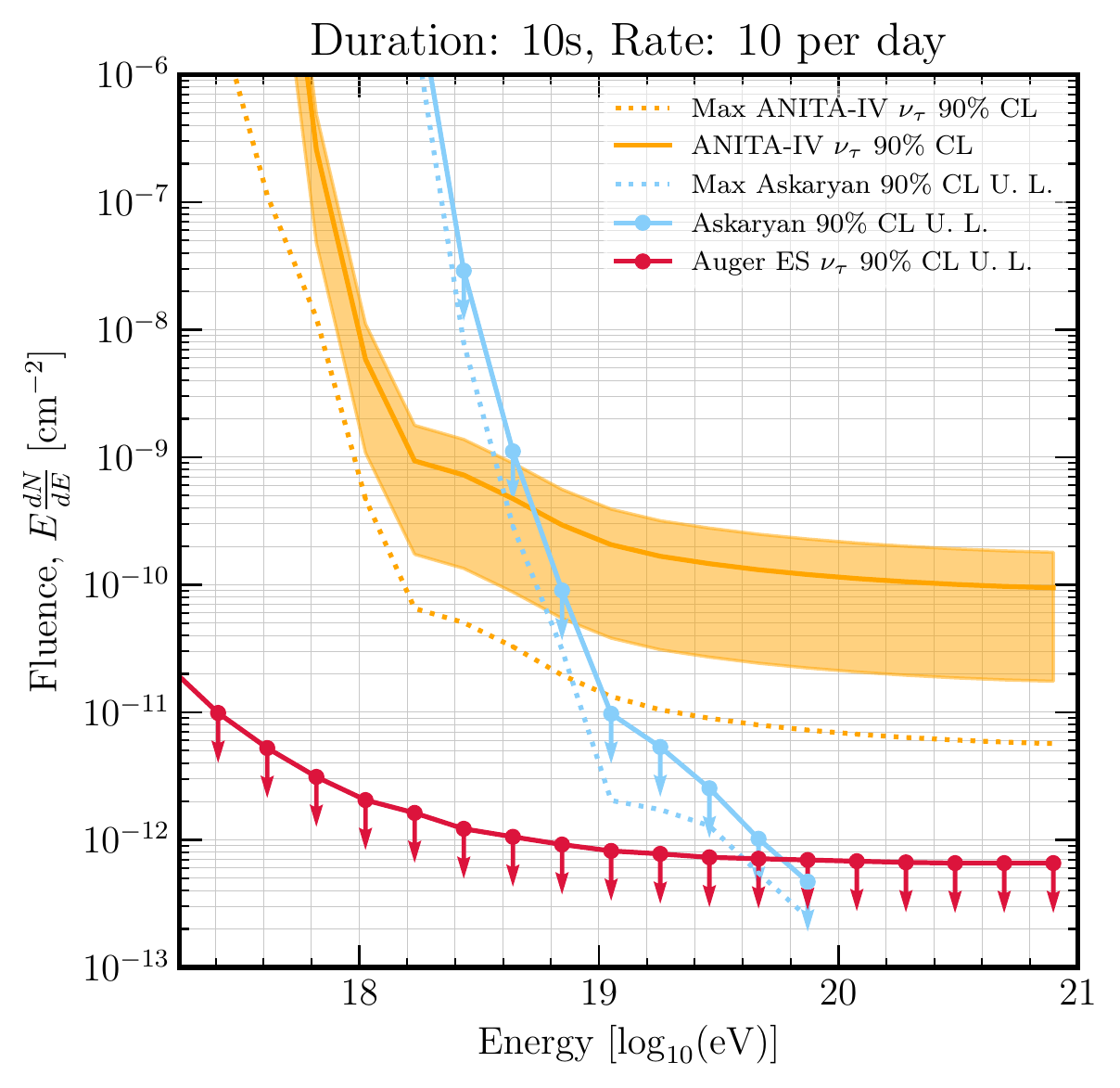}
    \includegraphics[width=\columnwidth]{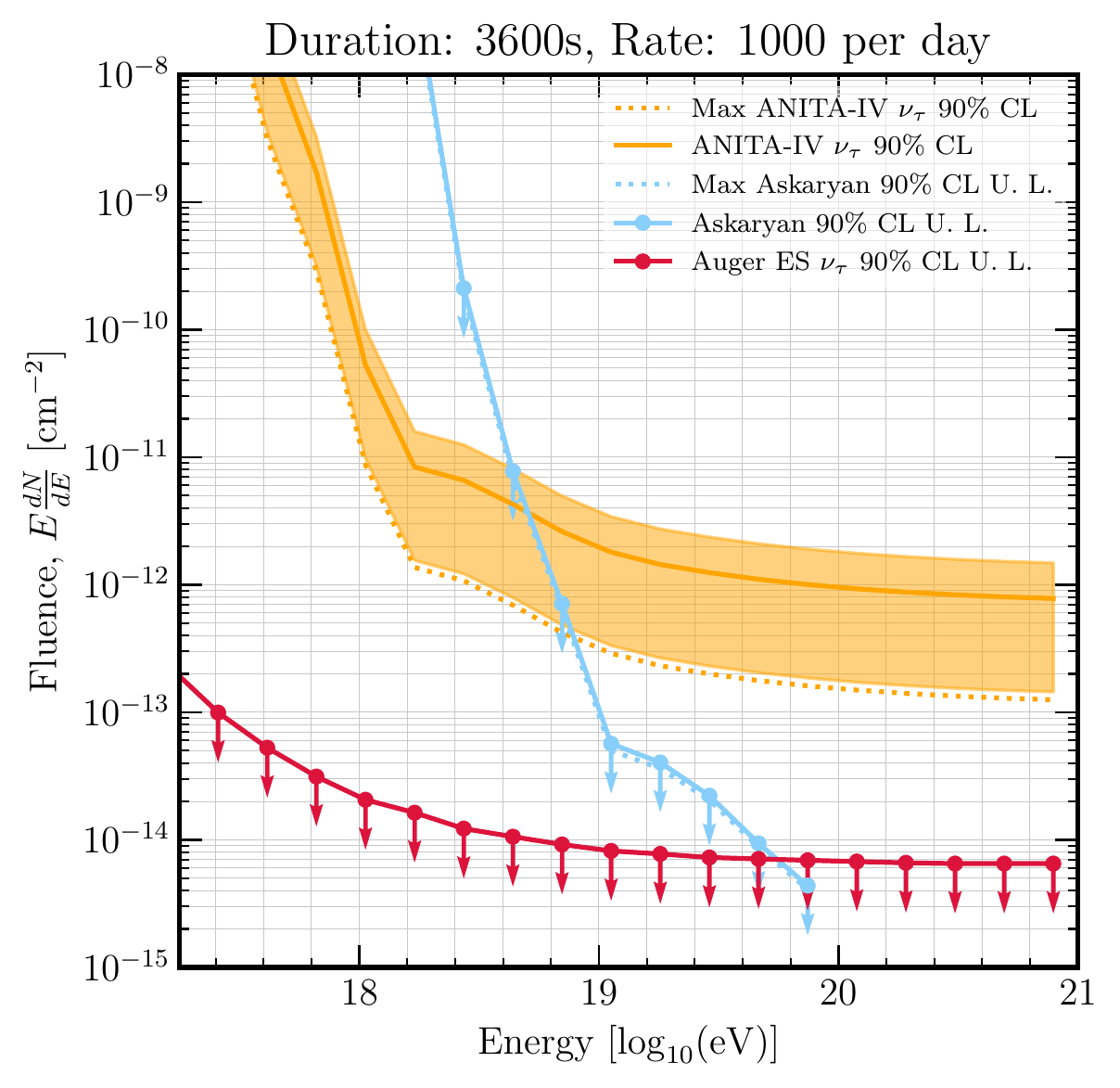}
    \caption{The fluence limits set by Auger $\nu_\tau$ and the ANITA-IV Askaryan channel assuming ANITA-IV observed $\sim3$ $\nu_\tau$ events via the EAS channel. For both transient source duration and full-sky event rates simulated here, the range of
    $\nu_\tau$ fluences consistent with our observation (orange band) are inconsistent with the limits set by Auger (red curve) across the entire simulated energy range and inconsistent with ANITA-IV's Askaryan limit
    (blue curve) above~$\sim10^{18}$~eV. All Auger results were calculated using the published effective areas from~\cite{2019JCAP...11..004A}.}
    \label{fig:model_independent_comparison}
  \end{figure}

We use a dedicated Monte Carlo simulation to calculate the distribution of possible outcomes for the number of detected events for the ANITA $\nu_\tau$,
ANITA Askaryan, and Auger channels. For a given transient duration $\Delta T$ and average full-sky event rate $r$,
we throw $N\sim\mathrm{Poisson}(rT_{\mathrm{auger}})$ random sources on the sky throughout the 
$\sim10$ years. For each source, we place a box-car (rectangular) time-dependent
flux model at the time of each simulated event with the given transient duration, $\Delta T$. We then calculate the total integrated exposure to each
of these transients using ANITA-IV's $\nu_\tau$ effective area (this work), ANITA-IV's Askaryan
point source effective area~\cite{2021JCAP...04..017A},
and Auger's upgoing $\nu_\tau$ effective area~\cite{2019JCAP...11..004A}. While Auger has sensitivity to downgoing
$\nu_\tau$ via in-air neutrino showers, this is significantly subdominant to the Earth-skimming $\nu_\tau$ channel and is 
therefore not included in this comparison. We calculate
the total sensitivity across all sources visible by each experiment assuming that the underlying
flux results in ANITA-IV observing $N_{\mathrm{true}}$ events, where $N_{\mathrm{true}}$ is sampled from Figure~\ref{fig:true_number_of_events} and is typically $\sim3$. Given
$N_{\mathrm{true}}$ detections by ANITA-IV's $\nu_\tau$ channel for this particular distribution of sources, we calculate corresponding limits on the fluence that would be set by ANITA-IV's Askaryan channel and Auger assuming that no events were detected in either observatory. This process is a single \textit{realization} of
ANITA-IV/Auger in the Monte Carlo simulation. This is repeated many times ($N_{\mathrm{src}} \in \left[10^{5}, 10^{6}\right]$) to accurately sample the distribution of possible transient
limits that each respective experiment may set. For a given underlying full-sky transient rate and duration, this Monte Carlo simulation accounts for fluctuations in the number and location of sources on the sky. The model independent limits on the fluence, calculated using this Monte Carlo, is shown in Figure~\ref{fig:model_independent_comparison} for two representative transient durations and rates.

  
For all simulated transient durations and full-sky rates, the observation of $\sim3$ $\nu_\tau$ events
is in strong tension with Auger across the full simulated energy range, and is also in tension with ANITA-IV's
Askaryan channel above $\sim10^{18.8}$~eV. This tension with ANITA's Askaryan channel could potentially
be resolved by a cut-off in the neutrino energy spectrum at or around $\sim10^{19}$~eV but this would
not eliminate the tension with Auger. As shown in Figure~\ref{fig:field_of_view}, ANITA-IV and Auger's Earth-skimming $\nu_\tau$ channels
have very similar instantaneous field-of-views so Auger immediately sets a stronger limit due to its $\sim120\times$ longer livetime
and larger effective area at lower energies (Figure~\ref{fig:peak_effective_area}).  The strength of the fluence limit set by each detector can vary significantly with the transient duration
and full-sky transient rate, but
Auger always sets a stronger limit than the ANITA-IV $\nu_\tau$ channel by an approximately constant factor due to the similarity of each observatories' fields-of-view and effective area. Above $10^{20}$~eV, ANITA-IV's Askaryan channel is able to set a stronger limit (in $\sim28$ days) than Auger for all simulated transient durations and full-sky transient rates.


\section{Discussion}
\label{sec:discussion}

The anomalous events found in ANITA-IV are significantly different from those found in the first and third flights in that they are closer to the horizon rather than being steeply upgoing. While this makes them more consistent with a Standard Model $\nu_\tau$  hypothesis as ANITA is maximally sensitive to $\nu_\tau$'s 
in the region below the horizon, the limits imposed by other observatories makes this an unlikely explanation.
For an in-depth discussion of backgrounds associated with these events, including ice surface and subsurface features, 
coherent backscattering, stopping radiation, and other effects, see the appendix of~\cite{2021PhRvL.126g1103G}.

In this section we briefly discuss the potential origin  of the low-frequency spectral discrepancies observed in 
several of the anomalous events and speculate on future investigations related to these events. 

\subsection{Potential Origin for the Low-Frequency Attenuation} 
\label{sec:low_frequency_anomaly}

In Figure~\ref{fig:deconvolved_waveforms}, we showed that while the spectra of events 4098827 and 72164985 match the expected exponential distribution, events 19848917 and 50549772 show attenuation at frequencies below $\sim500$~MHz. We explore several possibilities for the origin of this low-frequency (LF)
attenuation in two of the events: (1) geometries where ANITA simultaneously observes direct emission where an off-cone reflected emission whose inverted (from the reflection)
waveform interferes with the direct pulse; and (2) atmospheric propagation effects,
in particular tropospheric ducting, during the $\sim$600~km near-horizontal
propagation of the electric field from the decay point to ANITA.

The ``interfering'' reflected hypothesis (1) can be further broken into two
classes.  In both classes, the direct radio signal from an Earth-skimming air shower is observed along with a reflected signal. While they both propagate close to the horizon, the two classes are distinguished by the particles that generate the air shower and their incoming angle. In class (a), a cosmic ray produces an elongated air shower above the horizon, in the stratosphere. In class (b), an upgoing $\tau$ lepton decay produces a skimming air shower originating from below the horizon.  We consider both hypotheses by adding an inverted pulse -- representing the reflected signal -- to a non-inverted pulse -- representing the direct signal. The inverted and non-inverted pulses are delayed and summed in the time-domain to match the observed time-domain waveforms.

To recreate the observed waveforms,
the delay between the direct and reflected pulse must be less than $\sim$1~ns, or
$\lesssim30$~cm of total path length. An above-horizon cosmic ray geometry that allows for an off-axis
reflection detectable by ANITA with a $\lesssim$30~cm path length must propagate extremely close to the ground and is therefore strongly suppressed by the 
near horizon radio propagation effects discussed in
\cite{2020arXiv200805690A}. 

Furthermore, for each of the two events attenuated at low frequencies,
the emergence angle at the Earth is $2^{\circ}-3^{\circ}$. For a $\tau$ energy
of $\mathcal{O}(1-10\ \mathrm{EeV})$, the average $\tau$ lepton decay point is several kilometers
above the ground (the decay length for a 1~EeV $\tau$ is 47~km). With a $\lesssim30$~cm path length constraint, the
relatively high-altitude of the $\tau$ decay rules out any possible geometry for
a reflection. 

The reflected signal must also be of similar strength
to the direct pulse to create the observed waveforms shown in Fig.~
\ref{fig:deconvolved_waveforms}. As shown in \cite{Schoorlemmer:2015afa}, the total
reflection coefficient including Fresnel, roughness, and curvature effects,
approaches zero near the horizon significantly suppressing the strength of any
reflection, further disfavoring any reflected explanation for the observed
spectral attenuation below 250~MHz.



\subsection{Limitations of simulations}
\label{sec:simulation_challenges}

The air shower simulation code used in this work,
AIRES~\cite{1999astro.ph.11331S}, makes several assumptions that may affect the
simulation of near-horizontal cosmic ray showers from near-to or over-the
horizon. In particular, AIRES, as well as the other major EAS simulation code
CoREAS~\cite{1998cmcc.book.....H}: a) do not accurately simulate the
curvature of the Earth and atmosphere at the 600~km scale of the ANITA events
and therefore do not allow for ``over the horizon'' propagation; b) ignore the
refraction of the radio emission from the shower during propagation to the
receiving antennas; and c) use a \textit{geometric optics} formalism that
ignores any wave-like (diffraction, dispersion, ducting, etc.) effects that may
occurs in real atmospheres and alter the propagation of the radio emission from
the shower to the antenna. All of these effects are most dominant for events
originating near the horizon (i.e. propagating close to the surface) where the
Earth's curvature is most significant and many wave-like effects are possible
(i.e. tropospheric ducting, Fresnel zone attenuation or diffraction, etc.) which
are often strongly frequency-dependent and could potentially explain the anomalous low-frequency spectra observed in two of these ANITA events.

The authors of ZHAireS have investigated the effect of ray curvature for
highly-inclined \textit{reflected} UHECR showers (as might be seen by ANITA) and
found that for showers with zenith angles of $85^{\circ}$, the straight-ray
approximation was valid up to $\sim900$~MHz above which several changes in the
angular spectrum could be observed~\cite{Reflected_UHECR}. The four events
discussed in this work were observed at frequencies well below this frequency.
Furthermore, the $\tau$-induced EAS events visible by ANITA are also less likely
to be affected by these effects, compared to a high-zenith angle reflected EAS,
since a $\tau$ lepton at these energies typically decays tens or hundreds of
kilometers after leaving the Earth and therefore the shower typically develops
far from the horizon and potentially several kilometers above the surface, away
from the regions that are most affected by these approximations.

While this analysis has been performed using these existing tools as they are the best available at the time of writing, future efforts may help alleviate
some (but not all) of these issues. The upcoming next-generation shower simulation tool CORSIKA~8~\cite{corsika8} allows for simulating showers
in arbitrary 3D geometries so will allow for a correct treatment of Earth- and atmospheric curvature near the horizon. However, the current simulation programs of radio pulses in EAS based on superposition of contributions
from particle sub-tracks (ZHS~\cite{ZHS} and CoREAS~\cite{endpoint}) do not currently account for the geometric refraction of rays during propagation. 
Future versions of these packages such as those planned to be implemented in CORSIKA~8 may be able to incorporate these effects and may significantly alter these results. 

However, incorporating full-wave optic effects is a computationally challenging problem that will require the development of significantly new tools. The
standard high-fidelity full-wave electromagnetics simulation tool is the finite-difference time-domain (FDTD) algorithm but this requires extremely large
amounts of memory when simulating large volumes at high frequencies and simulating the propagation of ANITA's near horizon events is currently computationally intractable on even the world's largest supercomputers. Alternative methods, such as parabolic equation (PE) propagation, have the potential to provide
more accurate wave-like simulations for EAS than current tools (ZHAireS, CoREAS) but are still computationally expensive and have so far primarily been developed for defense-related radar propagation and are only beginning to be employed for ultrahigh-energy neutrino physics~\cite{parabolic}.



\subsection{Future Observations}
Several current and future experiments are designed to search for upgoing tau neutrinos via the $\tau$-induced air shower channel. PUEO, the follow-up mission to ANITA, has significantly improved sensitivity to $\tau$-induced EAS events and includes dedicated hardware to improve analysis and reduce backgrounds~\cite{Allison:2020emr}. Experiments searching for air showers from high elevation mountains~\cite{Nam:2016cib,Wissel:2020sec,Otte:2019knb,Alvarez-Muniz:2018bhp,Brown:2021tf}, balloons~\cite{Allison:2020emr}, and satellites~\cite{Olinto:2020oky} are also sensitive to events with similar geometries and origins. Given that these events are challenging to interpret under a $\nu_\tau$ hypothesis, it will be important to follow-up the ANITA observations in different locations (overlooking water, rock,) and from different altitudes (mountain, balloon, satellite) that will have different systematics and backgrounds.

The point-like $\nu_\tau$ analysis by Auger used for this work also only includes the surface detectors. A tau search using Auger's fluorescence detectors was also performed but only considered events with exit elevation angles greater than $20^\circ$ and so cannot be used to constrain these new ANITA-IV events~\cite{Abreu:2021AF,Caracas:2021Ia}.

\section{Conclusion}
\label{sec:conclusion}

We have analyzed the plausibility that the upgoing near-horizon ANITA-IV events are explained by $\tau$-lepton extensive air showers from skimming $\nu_\tau$ interactions in the Earth. To achieve this, we have applied detailed models of the $\nu_\tau\to\tau$ propagation through the Earth, radio emission from air showers, and the ANITA-IV detector. We have found consistency in the elevation angles and radio-frequency impulsive signatures of these events, namely the polarity and spectral shape of the events, with reconstructed $\nu_\tau$ energies in the 1 - 50~EeV range (depending upon assumptions regarding the underlying neutrino flux shape). We find that while these events are not observationally \textit{inconsistent} with UHE $\nu_\tau$'s, the implied fluence necessary for ANITA-IV to have observed $\sim3$ of these events is in tension with Auger's existing $\nu_\tau$ limits at all simulated energies and is also in tension with ANITA's Askaryan channel
above $10^{18.8}$~eV.

\begin{acknowledgments}
  We are especially grateful to the staff of the Columbia Scientific Balloon
  Facility for their generous support. We would like to thank NASA, the National
  Science Foundation, and those who dedicate their careers to making our science
  possible in Antarctica. This work was supported by the Kavli Institute for
  Cosmological Physics at the University of Chicago, the University of Hawai'i M\=anoa,
  and NASA bridge funding for ANITA-IV and ANITA's successor, PUEO. S. Wissel
  would also like to thank the California Polytechnic State University Frost Fund.
  This work has received financial support from Xunta de Galicia (Centro singular de investigaci\'on de Galicia accreditation 2019-2022), by European Union ERDF, by the ''Mar\'\i a de Maeztu'' Units of Excellence program MDM-2016-0692,
the Spanish Research State Agency and from Ministerio de Ciencia e Innovaci\'on PID2019-105544GB-I00
and RED2018-102661-T (RENATA).
  Computing resources were
  provided by the Research Computing Center at the University of Chicago and the
  Ohio Supercomputing Center at The Ohio State University. A. Connolly would
  like to thank the National Science Foundation for their support through CAREER
  award 1255557. The University College London group was also supported by the
  Leverhulme Trust. The National Taiwan University group is supported by
  Taiwan’s Ministry of Science and Technology (MOST) under its Vanguard Program
  106-2119-M-002-011. The following open-source tools were used in the
  preparation of this paper: \texttt{corner}~\cite{corner},
  \texttt{emcee}~\cite{emcee}, NumPy~\cite{numpy}, Matplotlib~\cite{matplotlib},
  Scipy~\cite{scipy}, and WebPlotDigitizer~\cite{Rohatgi2020}.
\end{acknowledgments}

\bibliographystyle{apsrev}
\bibliography{taunearhorizon}

\appendix

\end{document}

%% file: anita_revtex_institutes.tex

\newcommand{\atHarvard}{\affiliation{Harvard University, Cambridge, MA 02138.}}
\newcommand{\atUH}{\affiliation{Dept. of Physics and Astronomy, Univ. of Hawai'i, M\=anoa, HI 96822.}}
\newcommand{\atOSU}{\affiliation{Dept. of Physics, Center for Cosmology and AstroParticle Physics, Ohio State Univ., Columbus, OH 43210.}}
\newcommand{\atUCL}{\affiliation{Dept. of Physics and Astronomy, University College London, London, United Kingdom.}}
\newcommand{\atJPL}{\affiliation{Jet Propulsion Laboratory, California Institute for Technology,  Pasadena, CA 91109.}}
\newcommand{\atKU}{\affiliation{Dept. of Physics and Astronomy, Univ. of Kansas, Lawrence, KS 66045.}}
\newcommand{\atMoscow}{\affiliation{Moscow Engineering Physics Institute, Moscow, Russia.}}
\newcommand{\atWU}{\affiliation{Dept. of Physics, McDonnell Center for the Space Sciences, Washington Univ. in St. Louis, MO 63130.}}
\newcommand{\atUD}{\affiliation{Dept. of Physics, Univ. of Delaware, Newark, DE 19716.}}
\newcommand{\atNTU}{\affiliation{Dept. of Physics, Grad. Inst. of Astrophys., Leung Center for Cosmology and Particle Astrophysics, National Taiwan University, Taipei, Taiwan.}}
\newcommand{\atUC}{\affiliation{Dept. of Physics, Enrico Fermi Inst., Kavli Inst. for Cosmological Physics, Univ. of Chicago, Chicago, IL 60637.}}
\newcommand{\atSLAC}{\affiliation{SLAC National Accelerator Laboratory, Menlo Park, CA, 94025.}}
\newcommand{\atUCSD}{\affiliation{Center for Astrophysics and Space Sciences, Univ. of California, San Diego, La Jolla, CA 92093.}}
\newcommand{\atUCLA}{\affiliation{Dept. of Physics and Astronomy, Univ. of California, Los Angeles, Los Angeles, CA 90095.}}
\newcommand{\atCalPoly}{\affiliation{Physics Dept., California Polytechnic State Univ., San Luis Obispo, CA 93407.}}
\newcommand{\atPennState}{\affiliation{Dept. of Physics, Dept. of Astronomy \& Astrophysics, Pennsylvania State University, University Park, PA, 16802.}}

\newcommand{\atSantiago}{\affiliation{Instituto Galego de F\'isica de Altas Enerx\'ias, Universidade de Santiago de Compostela, Santiago de Compostela, Spain}}
\newcommand{\atSaoPaulo}{\affiliation{Universidade de S\~ao Paulo, Instituto de F\'isica, S\~{a}o Paulo, SP, Brazil}}
\newcommand{\atRadboud}{\affiliation{IMAPP, Radboud University Nijmegen, Nijmegen, The Netherlands}}

%% file: anita_revtex_authors.tex

 \author{R.~Prechelt}\atUH
 \author{S.~A.~Wissel}\atPennState\atCalPoly
 \author{A.~Romero-Wolf}\atJPL
 \author{C.~Burch}\atHarvard
 \author{P.~W.~Gorham}\atUH
 \author{P.~Allison}\atOSU
 \author{J.~Alvarez-Mu{\~n}iz}\atSantiago
 \author{O.~Banerjee}\atOSU
 \author{L.~Batten}\atUCL
 \author{J.~J.~Beatty}\atOSU
 \author{K.~Belov}\atJPL
 \author{D.~Z.~Besson}\atKU\atMoscow
 \author{W.~R.~Binns}\atWU
 \author{V.~Bugaev}\atWU
 \author{P.~Cao}\atUD
 \author{W.~Carvalho~Jr.}\atRadboud
 \author{C.~H.~Chen}\atNTU
 \author{P.~Chen}\atNTU
 \author{Y.~Chen}\atNTU
 \author{J.~M.~Clem}\atUD
 \author{A.~Connolly}\atOSU
 \author{L.~Cremonesi}\atUCL
 \author{B.~Dailey}\atOSU
 \author{C.~Deaconu}\atUC
 \author{P.~F.~Dowkontt}\atWU
 \author{B.~D.~Fox}\atUH
 \author{J.~W.~H.~Gordon}\atOSU
 \author{C.~Hast}\atSLAC
 \author{B.~Hill}\atUH
 \author{S.~Y.~Hsu}\atNTU
 \author{J.~J.~Huang}\atNTU
 \author{K.~Hughes}\atUC\atOSU
 \author{R.~Hupe}\atOSU
 \author{M.~H.~Israel}\atWU
 \author{K.~M.~Liewer}\atJPL
 \author{T.~C.~Liu}\atNTU
 \author{A.~B.~Ludwig}\atUC
 \author{L.~Macchiarulo}\atUH
 \author{S.~Matsuno}\atUH
 \author{K.~McBride}\atOSU
 \author{C.~Miki}\atUH
 \author{K.~Mulrey}\atUD
 \author{J.~Nam}\atNTU
 \author{C.~Naudet}\atJPL
 \author{R.~J.~Nichol}\atUCL
 \author{A.~Novikov}\atKU
 \author{E.~Oberla}\atUC
 \author{S.~Prohira}\atOSU\atKU
 \author{B.~F.~Rauch}\atWU
 \author{J.~Ripa}\atNTU
 \author{J.~M.~Roberts}\atUH\atUCSD
 \author{B.~Rotter}\atUH
 \author{J.~W.~Russell}\atUH
 \author{D.~Saltzberg}\atUCLA
 \author{D.~Seckel}\atUD
 \author{H.~Schoorlemmer}\atUH
 \author{J.~Shiao}\atNTU
 \author{S.~Stafford}\atOSU
 \author{J.~Stockham}\atKU
 \author{M.~Stockham}\atKU
 \author{B.~Strutt}\atUCLA
 \author{M.~S.~Sutherland}\atOSU
 \author{G.~S.~Varner}\atUH
 \author{A.~G.~Vieregg}\atUC
 \author{N.~Wang}\atUCLA
 \author{S.~H.~Wang}\atNTU
 \author{E.~Zas}\atSantiago
 \author{A.~Zeolla}\atPennState
\collaboration{ANITA Collaboration}\noaffiliation